\documentclass[10pt]{article}
\usepackage{arxiv}
\usepackage[utf8]{inputenc} 
\usepackage[T1]{fontenc}    
\usepackage{graphicx}
\usepackage[mathlines]{lineno}
\usepackage{amsmath,amssymb,amsthm}
\usepackage{mathtools}
\usepackage{xspace}
\usepackage{epsfig}
\usepackage[lined,noend,linesnumbered,ruled,titlenotnumbered]{algorithm2e}
\usepackage{color}
\usepackage{url}
\usepackage[square,numbers,compress,sort]{natbib}
\usepackage{tikz}
\usepackage{float}
\usepackage{booktabs}
\usepackage{makecell}
\usepackage{multicol}
\usepackage{multirow}
\usepackage{longtable}
\usepackage{array}
\usepackage[libertine,vvarbb]{newtxmath}
\usepackage{bm}
\usepackage{colortbl,soul,xcolor,bbm}
\usepackage{subcaption}
\usepackage{nicefrac}
\usepackage{microtype}      
\usepackage[inline]{enumitem}

\DeclareMathOperator*{\argmax}{\arg\!\max}
\DeclareMathOperator*{\argmin}{\arg\!\min}

\setlist[itemize]{align=parleft,left=0pt..1em}
\SetKwInput{Input}{Inputs}
\SetKwComment{tcc}{//}{}

\SetCommentSty{mycommfont}
\newtheorem{definition}{Definition}
\newtheorem{observation}{Observation}

\newcommand{\ignore}[1]{}
\usepackage[createShortEnv,conf={external appendix}]{proof-at-the-end}
\pratendSetGlobal{text link={}}

\begin{document}
\author{
Anh Viet Do \and
Mingyu Guo \and
Aneta Neumann \And
Frank Neumann
\affiliations
Optimisation and Logistics, School of Computer and Mathematical Sciences, The University of Adelaide
}
\title{Diverse Approximations for Monotone Submodular Maximization Problems with a Matroid Constraint}
\renewcommand{\shorttitle}{Diverse Approx. for Monotone Submodular Maximization Problems over a Matroid}
\date{}
\maketitle

\begin{abstract}
Finding diverse solutions to optimization problems has been of practical interest for several decades, and recently enjoyed increasing attention in research. While submodular optimization has been rigorously studied in many fields, its diverse solutions extension has not. In this study, we consider the most basic variants of submodular optimization, and propose two simple greedy algorithms, which are known to be effective at maximizing monotone submodular functions. These are equipped with parameters that control the trade-off between objective and diversity. Our theoretical contribution shows their approximation guarantees in both objective value and diversity, as functions of their respective parameters. Our experimental investigation with maximum vertex coverage instances demonstrates their empirical differences in terms of objective-diversity trade-offs. 
\end{abstract}
\section{Introduction}
Optimization research has seen rising interest in diverse solutions problems, where multiple maximally distinct solutions of high quality are sought instead of a single solution \cite{Ingmar2020,Baste2020,hanaka2020finding,Fomin2021,Fomin20211,hanaka2021,hanaka2022}. This class of problem is motivated by practical issues largely overlooked in traditional optimization. Having diverse solutions gives resilient backups in response to changes in the problems rendering the current solution undesirable. It also gives the users the flexibility to correct for gaps between the problem models and real-world settings, typically caused by estimation errors, or aspects of the problem that cannot be formulated precisely \cite{Schittekat2009}. Furthermore, diverse solution sets contain rich information about the problem instance by virtue of being diverse, which helps augment decision making capabilities. While there are methods to enumerate high quality solutions, having too many overwhelms the decision makers \cite{Glover2000}, and a small, diverse subset can be more useful. It is also known that k-best enumeration tends to yield highly similar solutions, motivating the use of diversification mechanisms \cite{Wang2013,Yuan2015,Hao2020}.

The diverse solutions problem have been studied as an extension to many important and difficult problems. Some examples of fundamental problems include constraint satisfaction and optimization problems \cite{Emmanuel05,Petit15,Ruffini2019}, SAT and answer set problem \cite{Nadel2011,Eiter2009}, and mixed integer programming paradigms \cite{Glover2000,Danna,Trapp2015}. More recently, the first provably fixed-parameter tractable algorithms have been proposed for diverse solutions to a number of graph-based vertex problems \cite{Baste2020}, as motivated by the complexity of finding multiple high performing solutions. This inspired subsequent research on other combinatorial structures such as trees, paths \cite{hanaka2020finding,hanaka2021}, matching \cite{Fomin20211}, independent sets \cite{Fomin2021}, and linear orders \cite{Arrighi2021}. Furthermore, general frameworks have been proposed for diverse solutions to any combinatorial problem \cite{Ingmar2020,hanaka2022}. To address the need to obtain both quality and diversity, multicriteria optimization has been considered, leading to interesting results \cite{Gao2022}. These are mostly applied to problems with linear objective functions and specific matroid intersection constraints.

In this work, we are interested in diverse solutions problem in the domain of submodular optimization, which has been enjoying widespread interests. It captures the diminishing returns property that arises in many real-world problems in machine learning, signal processing \cite{Tohidi2020}, sensor placement \cite{Krause2005}, data summarization \cite{Lin2011,Mirzasoleiman2013}, influence maximization \cite{Kempe2015}, to name a few. Moreover, its hardness (as it generalizes many fundamental NP-hard combinatorial problems) and well-structuredness (which facilitates meaningful results \cite{Vondrk2013}) mean the problem class also sees much attention from theoretical perspectives, leading to interesting insights \cite{Nemhauser1978,Fisher1978,Calinescu2011,Krause,Chekuri2014}. It is important to distinguish between the diverse solutions extension to submodular optimization and results diversification \cite{Zheng2017}, the latter of which considers diversity as a measure of a solution (i.e. a selection of results) and optimizes it along with a submodular utility function.

\paragraph{Our Contributions} We investigate the problem of finding a given number of diverse solutions to maximizing a monotone submodular function over a matroid, with a lower bound on solutions' objective values. Matroids are a type of independence system that can be used to model constraints in many important problems, and have been studied in in submodular optimization literature \cite{Conforti1984,Lee2010,Calinescu2011,Kashaev2011,Chekuri2014}, even recently appeared in diverse solutions research \cite{Fomin2021}. Among them, uniform matroids which characterize cardinality constraints, and their extension, partition matroids, are often considered in budgeted optimization (e.g. \cite{lin-bilmes-2010-multi}). We consider the distance-sum measure of diversity, which is often chosen for diverse solutions problems \cite{hanaka2020finding,Baste2020,hanaka2021,hanaka2022,Gao2022}. Its sole reliance on the ground set elements' representation in the solution set implies generalizability to other diversity measures such as entropy. Our contributions are as follows:
\begin{itemize}
\item We propose two simple greedy algorithms which are suitable to deal with the objective requirement, as greedy algorithms are known to perform well on monotone submodular maximization \cite{Nemhauser1978,Fisher1978}. The novelty lies in the additional parameters, which adjust the trade-off between guarantees on objective values and diversity. We position our algorithms as simpler, zeroth-order (in terms of objective and independence oracles) alternatives to general frameworks for diverse solutions in recent literature, which have not been analyzed in submodular optimization context.
\item We provide analyses of these algorithms in terms of their objective-diversity guarantees trade-offs. Our results are formulated as functions of their respective parameters, thus giving a general guidance on parameter selection. We also give sharpened bounds for cases with uniform matroids, as motivated by the prevalence of cardinality constraints. From these results, we point out settings that guarantee constant approximation ratios in objective, diversity, or both. Our tightness constructions also indicate certain features of matroids that make them pathological to these algorithms.
\item We carry out an experimental investigation with maximum vertex coverage instances subjected to uniform and partition matroid constraints, to observe the algorithms' empirical performances in exhaustive parameter settings. The results indicate that while both algorithms produce nearly optimal solutions with reasonable diversity in many parameter settings, the simpler of the two actually provides better objective-diversity trade-offs across all problem settings. Additionally, these establish an empirical baseline for the diverse solutions problem considered in this work.
\end{itemize}
\section{Preliminaries}\label{sec:prelim}
In this section, we present the problem and relevant definitions, and give some observations that are helpful in our analyses.
\subsection{Problem and Definitions}
A multiset is a collection that can contain duplicates (e.g. $\{1,1,2\}$). For a set $A$, we denote the collection of multisets of elements in $A$ with $A^*$, and $A^r\subseteq A^*$ contains $r$-size\footnote{In this work, we use ``$r$-size'' to mean ``containing $r$ elements''.} multisets for some integer $r$. The problem we investigate is as follows: given integer $r\geq2$, $\alpha\in[0,1]$, a $(f,S,d,r,\alpha)$-instance asks for a multiset\footnote{Satisfying self-avoiding constraint requires algorithmic treatment beyond this work's scope.} of solutions in
\begin{equation}\label{eq:problem}
\argmax_{P\in S^r}\left\lbrace d(P):\forall x\in P,f(x)\geq\alpha\max_{y\in S}f(y)\right\rbrace.
\end{equation}
where the objective function, $f:2^V\to\mathbb{R}$ is non-negative\footnote{The non-negativity assumption is widely used in literature to ensure proper contexts for multiplicative approximation guarantees, which this work includes. This also applies to diversity w.l.o.g.} and non-decreasing submodular, $S=\mathcal{I}$ for some matroid $M=(V,\mathcal{I})$, and $d$ is a diversity measuring function defined over $(2^V)^*$. We do not consider non-increasing $f$ due to trivial instances where achieving any positive diversity\footnote{Assuming the diversity measure returns $0$ on duplicate-only multisets.} necessitates degrading solutions beyond the feasibility limit. As per standard practice, we use ``monotone'' to mean ``non-decreasing'' in this paper. We call a multiset $P$ feasible to the $(f,S,d,r,\alpha)$-instance if $P\in S^r$ and every solution in $P$ is a $\alpha$-approximation of $f$ over $S$, which is a solution $x\in S$ such that $f(x)\geq\alpha\max_{y\in S}f(y)$. We also briefly give relevant definitions and assumptions.

\begin{definition}
Function $f:2^V\to\mathbb{R}$ is monotone if $f(x)\leq f(y)$ for all $x\subseteq y\subseteq V$.
\end{definition}
\begin{definition}
Function $f:2^V\to\mathbb{R}$ is submodular if $\forall x,y\subseteq V,f(x)+f(y)\geq f(x\cup y)+f(x\cap y)$ or equivalently $\forall x\subseteq y\subseteq V,v\in V\setminus y,f(x\cup\{v\})-f(x)\geq f(y\cup\{v\})-f(y)$.
\end{definition}
For problem \eqref{eq:problem}, we assume w.l.o.g. that $f(\emptyset)=0$, since a multiset feasible to a $(f,S,d,r,\alpha)$-instance is also feasible to the $(f+f',S,d,r,\alpha)$-instance for some constant non-negative function $f'$. We assume for our problem that $f$ is given as a value oracle.

For matroid theory concepts, we adopt terminologies from the well-known text book \cite{Oxley2011} on the subject.
\begin{definition}
A tuple $M=(V,\mathcal{I}\subseteq2^V)$ is a matroid if \begin{enumerate*}[label=\itshape\alph*\upshape)]
\item $\emptyset\in\mathcal{I}$,
\item $\forall x\subseteq y\subseteq V, y\in\mathcal{I}\implies x\in\mathcal{I}$,
\item $\forall x,y\in\mathcal{I},\allowbreak|x|<|y|\implies\exists e\in y\setminus x,x\cup\{e\}\in\mathcal{I}$. The set $V$ is the ground set, and $\mathcal{I}$ is the independence collection. A base of $M$ is a maximal set in $\mathcal{I}$.
\end{enumerate*}
\end{definition}
\begin{definition}
Given a matroid $M=(V,\mathcal{I})$,
\begin{itemize}
\item the rank function of $M$, $r_M:2^V\to\mathbb{N}$, is defined as $r_M(x)=\max\{|y|:y\in 2^x\cap\mathcal{I}\}$, and the rank of $M$ is $r_M=r_M(V)$,
\item the closure function of $M$, $cl_M:2^V\to2^V$, is defined as $cl_M(x)=\{v\in V:r_M(x\cup\{v\})=r_M(x)\}$,
\item a loop of $M$ is a $v\in V$ such that $\{v\}\notin\mathcal{I}$.
\end{itemize}
\end{definition}
To give examples, a $K$-rank uniform matroid over $V$ admits the independence collection $\mathcal{I}=\{x\subset V:|x|\leq K\}$ which we denote with $\mathcal{U}_{V,K}$. A partition matroid admits the independence collection $\mathcal{I}=\{x\subset V:\forall i=1,\ldots,k,|x\cap B_i|\leq d_i\}$ for some partitioning $\{B_i\}_{i=1}^k$ of $V$ and their corresponding thresholds $\{d_i\}_{i=1}^k$. In graph theory, a graphic matroid $M=(E,\mathcal{I})$ defined over a undirected graph $G=(V,E)$ is such that $\mathcal{I}$ contains all edge sets $x$ where $G'=(V,x)$ has no cycle. A base of a graphic matroid is a spanning forest in the underlying graph, which itself is an object of much interest. Dual to the graphic matroid, the bond matroid $M^*=(E,\mathcal{I}^*)$ is such that $\mathcal{I}^*$ contains all edge sets $x$ where $G^*=(V,E\setminus x)$ has the same number of connected components as $G$.

For the problem \eqref{eq:problem}, we assume that $M$ is loop-free and $|V|\geq1$, implying $r_M>0$. It is known that rank functions are monotone submodular, and closure functions are monotone, i.e. $x\subseteq y\implies cl_M(x)\subseteq cl_M(y)$ \cite{Oxley2011}. We also assume that for a matroid, we are given an independence oracle answering whether a set is independent.

Finally, we consider the distance-sum diversity function, which is the usual choice in literature on diverse solutions problems \cite{hanaka2020finding,Baste2020,hanaka2021,hanaka2022,Gao2022}. The function is defined over multisets of solutions as $ss(P)=\sum_{x,y\in P}|x\Delta y|$ where $\Delta$ is the symmetric difference between two sets, and its size is the Hamming distance. To be precise, each pairwise distance is counted once in an evaluation of $ss$.

Under this setting, the problem \eqref{eq:problem} is equivalent to the dispersion problem over the ground set that is the collection of all $\alpha$-approximations of $f$ over $\mathcal{I}$. The dispersion problem is known to be NP-hard in the ground set's size, even with known ground sets and metric distance functions \cite{Wang1988,Erkut1990,Ravi1994,Chandra1996}; for our problem, the collection is neither known nor necessarily small. On the other hand, \cite{hanaka2022} showed that this problem admits a poly-time $\max\{1-2/r,1/2\}$-approximation scheme, predicated on a poly-time top-$r$ enumeration scheme over this collection maximizing $ss$. We are not aware of such a scheme for $\alpha$-approximations to submodular maximization over a matroid, and we recognize this as an interesting problem in its own right. That said, it is likely that algorithms resulted from this line of ideas will have significantly larger asymptotic run-times than those of the algorithms we present in this work.
\subsection{Some Useful Properties}
First, we observe that the value of $ss$ is related to the occurrences of each elements of $V$ in the multiset. Let $P$ be a $r$-size multiset of subsets of $V$, and for $i=1,\ldots,|V|$, $n_i(P)=|\{x\in P: i\in x\}|$, we have
\begin{equation}\label{eq:ss_counts}
ss(P)=\sum_{i\in V}n_i(P)[r-n_i(P)].
\end{equation}
This means the function can be decomposed into disjoint subsets of $V$: given a partitioning $\{V_i\}_{i=1}^k$ of $V$, we have $ss(P)=\sum_{i=1}^kss(\{x\cap V_i: x\in P\})$. This property can significantly simplify analyses.

We would also like to bound the maximum achievable diversity in various settings. While without the constraint on the values of $f$, this bound can be computed (over a matroid) exactly and efficiently, e.g. by using the method in \cite{hanaka2020finding}, estimating it with a formula can be useful. To this end, we define a function $g:\mathbb{N}^3\to\mathbb{N}$ with
\[g(a,b,c)=aq(c-q)+m(c-2q-1),\]
where $h=\min\{b,a/2\}$, and $m\in[0,a),q$ are integers such that $\lceil c/2\rceil\lceil h\rceil+\lfloor c/2\rfloor\lfloor h\rfloor=qa+m$. This function returns the maximum $ss$ values of a $c$-size multisets of at most $b$-size subsets of a $a$-size ground set (Theorem \ref{theorem:matroid_optimum_bound}). Also, we let $g(0,\cdot,\cdot)=g(\cdot,0,\cdot)=g(\cdot,\cdot,1)=0$. For convenience, let $\delta:\mathbb{N}^2\to\mathbb{N}$ be defined with $\delta(a,b)=a-2b-1$, we have
\begin{align*}
\forall x\in P,e\in V\setminus x,&ss(P\setminus\{x\}\cup\{x\cup\{e\}\})-ss(P)=|P|-2n_e(P)-1=\delta(|P|,n_e(P)).
\end{align*}
This expression exposes the connection between $g$ and the process of adding elements into solutions in $P$, which is relevant to the algorithms we consider in this work. That is, we can rewrite $g$ using $\delta$: $g(a,b,c)=a\sum_{i=0}^{q-1}\delta(c,i)+m\delta(c,q)$; this simplifies the proof of its monotonicity.

\begin{lemmaE}[][category=gmonotone,end]\label{lemma:g_monotone}
Function value $g(a,b,c)$ is monotone in $a$, $b$ and $c$.
\end{lemmaE}
\begin{proof}
Let $G(a,b,c)$ be the multiset of values of the summands $\delta$ in $g(a,b,c)$. For any $a'>a$, we have $|G(a,b,c)|\leq|G(a',b,c)|$ and a bijection $g':G(a,b,c)\to G'\subseteq G(a',b,c)$ such that $d\leq g'(d)$ for all $d\in G(a,b,c)$. This implies $g(a,b,c)\leq g(a',b,c)$. For any $b'>b$, we have $G(a,b,c)\subseteq G(a,b',c)$, so $g(a,b,c)\leq g(a,b',c)$. For any $c'>c$, we have $G(a,b,c)\subseteq G(a,b,c')$ and a bijection $g':G(a,b,c)\to G'\subseteq G(a,b,c')$ such that $d=g'(d)-c'+c$ for all $d\in G(a,b,c)$, so $g(a,b,c)<g(a,b,c')$.
\end{proof}

Here, we include an inequality which gives an intuitive bound of a result in Section \ref{sec:algorithm}.

\begin{lemmaE}[][category=gsumboundratio,end]\label{lemma:g_sum_bound_ratio}
Given integers $a,b,c\geq1$ and $k\geq0$, $g(\lceil ka/b\rceil,k,c)\geq  kg(a,b,c)/y$.
\end{lemmaE}
\begin{proof}
Let $G(a,b,c)$ be the multiset of values of the summands $\delta$ in $g(a,b,c)$, we have $|G(a,b,c)|\leq k|G(\lceil ka/b\rceil,k,c)|/y$. Let $S(a,b,c)$ be the set derived from $G(a,b,c)$ (i.e. all duplicates removed), we have $S(\lceil ka/b\rceil,k,c)\subseteq S(a,b,c)$ and $\forall a\in S(a,b,c)\setminus S(\lceil ka/b\rceil,k,c),a<\min\{b\in S(\lceil ka/b\rceil,k,c)\}$. Furthermore, every element in $S(\lceil ka/b\rceil,k,c)$ except the minimum has multiplicity in $G(\lceil ka/b\rceil,k,c)$ equals $\lceil ka/b\rceil/a$ times its multiplicity in $G(a,b,c)$. Therefore, the sum of elements in $G(\lceil ka/b\rceil,k,c)$ is at least $k/b$ the sum of elements in $G(a,b,c)$, hence the claim.
\end{proof}

To establish an upper bound on diversity, we use the following straightforward observation from the fact that uniform matroid constraints are the least restrictive.

\begin{observation}\label{observation:uniform_optimum_bound}
Given a set $V$, function $f$ over $2^V$, matroids $M=(V,\mathcal{I})$ and $M'=(V,\mathcal{I}')$ where $M$ is uniform and $r_M\geq r_{M'}$, then the optimal value for the $(f,\mathcal{I}',d,r,0)$-instance cannot exceed that for the $(f,\mathcal{I},d,r,0)$-instance with any $r\geq1$, and real function $d$ over $(2^V)^*$.
\end{observation}

With this, we can use uniform matroids to formulate a simple upper bound, which is also tight for some non-uniform matroids and, surprisingly, any value of the threshold ratio $\alpha$.

\begin{theoremE}[][category=matroidoptimumbound,end]\label{theorem:matroid_optimum_bound}
The optimal value for a $(f,\mathcal{I},ss,r,\alpha)$-instance for some matroid $M=(V,\mathcal{I})$, function $f$ over $2^V$, integer $r\geq1$, and $\alpha\in[0,1]$ is at most $g(|V|,r_M,r)$. Moreover, this bound is tight for any $|V|\geq1$, $r\geq1$, $\alpha\in[0,1]$, and matroid rank $r_M\in[1,|V|]$, even if the matroid is non-uniform.
\end{theoremE}
\begin{proof}
By Equation \eqref{eq:ss_counts}, $ss(P)$ is the sum of negative quadratic functions of $n_i(P)$. Each of these summands is maximized at $r/2$, so assuming $M$ is uniform and $\alpha=0$, $ss(P)$ is maximized if every solution in $P$ contains $h=\min\{r_M,|V|/2\}$ elements, since it cannot exceed $r_M$ elements. Therefore, the maximum $ss(P)$ under this assumption is $g(|V|,r_M,r)$. According to Observation \ref{observation:uniform_optimum_bound}, this assumption is ideal, so this bound cannot be exceeded by any feasible set $P$ under any matroid $M$ and $\alpha\in[0,1]$.

We prove tightness by construction. Given integers $n\geq1$, $s\in[1,n]$, $r\geq2$, let $x=(x_i)_{i=1}^n$ be the characteristic vector of subsets of $V$, $m\in[0,s),q$ be integers where $n=qs+m$, $f(x)=\sum_{k=0}^{q}\sum_{j=1}^{m}c_{j}x_{ks+j}$ where $c_i$ is a non-negative real for all $i=1,\ldots,m$, and a $s$-rank matroid $M=(V,\mathcal{I})$ where $\mathcal{I}=\{x\subseteq V:|x|\leq s\wedge \forall h\in[1,m],\sum_{i=0}^sx_{qi+h}\leq1\}$. We have $OPT=\max_{x\in\mathcal{I}}\{f(x)\}=\sum_{j=1}^mc_j$. Let $P$ be a $r$-size multiset $\left\lbrace\{i\}_{i=(j\mod q+1)s+1}^{(j\mod q+1)s+m}\cup\{i\}_{i=(j\mod q)s+m+1}^{(j\mod q)s+s}\right\rbrace_{j=1}^r$, we have $ss(P)=g(|V|,s,r)$ and for all $x\in P$, $f(x)=OPT$ and $x\in\mathcal{I}$. Thus, $P$ is optimal for any $\alpha\in[0,1]$.
\end{proof}

In augment-type algorithms like greedy, how the feasible selection pool for a partial solution (i.e. set of elements that can be added without violating constraints) changes over the course of the algorithm influences the guaranteed quality of the final output. This insight was made evident in seminal works on greedy algorithms \cite{Fisher1978,Nemhauser1978}, and is replicated in subsequent works on submodular optimization under more complex constraints. This is especially important in diverse solutions, as high diversity can be seen as additional restrictions on the selection pool. In the context of matroid constraint, this pool is determined by the partial solution's closure, thus we include an observation connecting closures to the upper bound on diversity.

\begin{lemmaE}[][category=matroidfreedomlimit,end]\label{lemma:matroid_freedom_limit}
Let $M=(V,\mathcal{I})$ be a matroid (may contain loops), and $x\in\mathcal{I}$, then for all $y\in\mathcal{I}$, $|y\cap cl_M(x)|\leq|x|$. By extension, $|y\cap z|\leq r_M(z)$ for all $z\subseteq V$.
\end{lemmaE}
\begin{proof}
This is clearly the case if $cl_M(x)=x$. Assuming otherwise, and there is $y\in\mathcal{I}$ where $|y\cap cl_M(x)|>|x|$, then by the exchange property between independent sets, there is $e\in y\cap cl_M(x)\setminus x$ where $x\cup\{e\}\in\mathcal{I}$. This implies $(y\cap cl_M(x)\setminus x)\nsubseteq cl_M(x)\setminus x$, a contradiction.
\end{proof}
Lemma \ref{lemma:matroid_freedom_limit} lets us sharpen the upper bound on $ss$ values for highly non-uniform matroids.
\begin{lemmaE}[][end]\label{lemma:optimum_bound_closure}
Given a matroid $M=(V,\mathcal{I})$ (may contain loops) and integer $r\geq1$, then for any $P\in\mathcal{I}^r$,
\begin{align*}
ss(P)\leq\min_{x\in\mathcal{I}}\{&g(|V|-|cl_M(x)|,r_M-\lfloor n_x\rfloor,r)+g(|cl_M(x)|,\lceil n_x\rceil,r)\},
\end{align*}
where $n_x=\min\{r_M|cl_M(x)|/|V|,|x|\}$. There exists a matroid where equality holds.
\end{lemmaE}
\begin{proof}
Let $P$ be a $r$-size multiset of independent sets in $M$, and $x\in\mathcal{I}$, then Lemma \ref{lemma:matroid_freedom_limit} and the properties of matroids imply that for all $y\in P$, $|y\cap cl_M(x)|\leq|x|$. This means $ss(P)\leq g(|cl_M(x)|,|x|,r)+g(|V|-|cl_M(x)|,r_M,r)$, which is the sum of maximum achievable $ss$ values within $cl_M(x)$ and $V\setminus cl_M(x)$, respectively. This bound can be sharpened by the fact that $|y|\leq r_M$ for all $y\in P$. Since $ss(P)$ is maximized when elements in $V$ are included in equal numbers of sets in $P$. This means assuming an ideal scenario where independence is not violated in any other way, $P$ maximizing $ss$ must minimize the gap between $\sum_{y\in P}|y\cap cl_M(x)|/|cl_M|$ and $\sum_{y\in P}|y\setminus cl_M(x)|/|V\setminus cl_M(x)|$. Since $g$ can be used to characterize maximum $ss$ in such a setting, we have any such $P$ must satisfy
\[ss(P)\leq g(|V|-|cl_M(x)|,r_M-\lfloor n_x\rfloor,r)+g(|cl_M(x)|,\lceil n_x\rceil,r),\]
where $n_x=\min\{r_M|cl_M(x)|/|V|,|x|\}$. Since this holds for any $x\in\mathcal{I}$, the claim follows.
\end{proof}

We give two more useful inequalities regarding function $g$.

\begin{figure}[ht!]
\centering
\includegraphics[width=.7\linewidth]{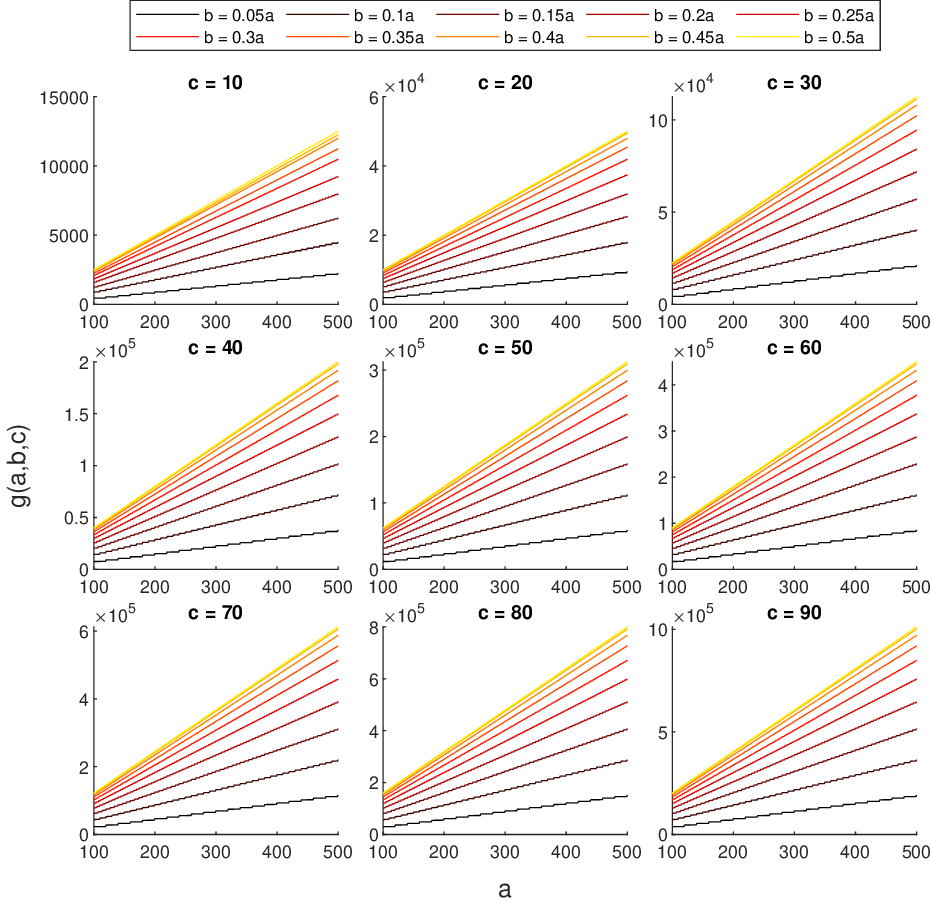}\vspace{-5pt}
\caption{Values of $g(a,b,c)$ with various input values where $b$ is assigned fixed fractions of $a$.}\label{fig:g}
\end{figure}

\begin{lemmaE}[][end]\label{lemma:g_sum_bound}
Let $c,m\geq1$, $\{a_i\}_{i=1}^m$, $\{b_i\}_{i=1}^m$ be non-negative integers, $g\left(\sum_{i=1}^ma_i,\sum_{i=1}^mb_i,c\right)\geq\sum_{i=1}^mg(a_i,b_i,c)$.
\end{lemmaE}
\begin{proof}
Let $\{U_i\}_{i=1}^m$ be disjoint sets where $|U_i|=a_i$ for $i=1,\ldots,m$, $V=\bigcup_{i=1}^mU_i$, $A=\left\lbrace x\subseteq V:|x|\leq\sum_{i=1}^mb_i\right\rbrace$, $B=\left\lbrace y\subseteq V:\forall i=1,\ldots,m,|y\cap U_i|\leq b_i\right\rbrace$. Define $A'$ and $B'$ as the collections of $c$-size multisets of sets in $A$ and $B$, respectively. Given the lack of any other constraint, Theorem \ref{theorem:matroid_optimum_bound} implies that $g\left(\sum_{i=1}^ma_i,\sum_{i=1}^mb_i,c\right)=\max_{P\in A'}ss(P)$. Using the decomposition of $ss$, we also have $\sum_{i=1}^mg(a_i,b_i,c)=\max_{P\in B'}ss(P)$. The claim follows from $A'\supseteq B'$, which is the case as $A\supseteq B$.
\end{proof}
\begin{lemmaE}[][end]\label{lemma:greedy_monotone_matroid_lowerbound_helper}
Given integers $a\geq2$, $c\geq1$, $b\in[1,a-1)$, $l\in[0,\lceil c/2\rceil)$, $m=\lceil c/2\rceil-l$, a non-increasing integer sequence $(a_i)_{i=1}^m$ such that $a_i\in[0,a]$ for all $i$, $a_1\geq b$, $bm-\sum_{i=1}^ma_i\leq l(a-b)-c$ and $l(a-b)+a_1-b\geq c$, then
\begin{align*}
a\sum_{i=1}^l\delta(c,i-1)+\sum_{i=l+1}^{l+m}\delta(c,i-1)a_{i-l}\geq g(b,b,c)+g(a-b,1,h)-h(h-c),
\end{align*}
where $h=l(a-b)+a_1-b+\min\left\lbrace\sum_{i=2}^ma_i-b(m-1),0\right\rbrace$.
\end{lemmaE}
\begin{proof}
Let $\Delta=a-b$, the left hand side be $L$, and $j\in[2,m]$ be such that $a_{j-1}\geq b$ and $a_j<b$, we split $L=L_1+L_2$ where
\begin{align*}
L_1=b\sum_{i=1}^{l+j-1}\delta(c,i-1)+\sum_{i=l+j}^{l+m}\delta(c,i-1)a_{i-l}\quad\text{and}\quad L_2=\Delta\sum_{i=1}^l\delta(c,i-1)+\sum_{i=l+1}^{l+j-1}\delta(c,i-1)(a_{i-l}-b).
\end{align*}
We have
\begin{align*}
L_1-g(b,b,c)=L_1-b\sum_{i=1}^{\lceil c/2\rceil}\delta(c,i-1)=L_1-b\sum_{i=1}^{l+m}\delta(c,i-1)=\sum_{i=l+j}^{l+m}\delta(c,i-1)(a_{i-l}-b).
\end{align*}
If $\sum_{i=2}^ma_i<b(m-1)$ then let $h'=\lfloor h/\Delta\rfloor$, $k=\max\{l,h'+1\}$ and $d\in[0,\Delta)$ such that $h\equiv d\mod\Delta$, we have
\begin{align*}
L_2-g(\Delta,1,h)&=L_2-\sum_{i=1}^{h'}\delta(h,i-1)-d\delta\left(h,h'\right)=L_2-\sum_{i=1}^{h'}\delta(c,i-1)-d\delta\left(c,h'\right)-h(h-c)\\
&=(\Delta-d)\delta\left(c,h'\right)+\Delta\sum_{i=h'+2}^{l}\delta(c,i-1)+\sum_{i=k+1}^{l+j-1}\delta(c,i-1)(a_{i-k}-b)-h(h-c)\\
&\geq\left[b(m-1)-\sum_{i=2}^{m}a_i\right]\delta(c,l+j-2)+\sum_{i=l+1}^{l+j-1}\delta(c,i-1)(a_{i-l}-b)-h(h-c)\\
&\geq\left[b(j-2)-\sum_{i=2}^{j-1}a_i\right]\delta(c,l+j-2)+\sum_{i=l+1}^{l+m}\delta(c,i-1)|a_{i-l}-b|-h(h-c)\\
&\geq\sum_{i=l+2}^{l+j-1}\delta(c,i-1)(b-a_{i-l})+\sum_{i=l+1}^{l+j-1}\delta(c,i-1)(a_{i-l}-b)-L_1+g(b,b,c)-h(h-c)\\
&\geq\delta(c,l)(a_1-b)-L_1+g(b,b,c)-h(h-c)\geq g(b,b,c)-L_1-h(h-c),
\end{align*}
where the inequalities follow from $\delta$ being decreasing in the second parameter, and $b-a_i$ being non-positive for all and only $i\in[1,j-1]$. Now, if $\sum_{i=2}^ma_i\geq b(m-1)$, then $\sum_{i=2}^{j-1}a_i-(j-2)b\geq (m-j+1)b-\sum_{i=j}^{m}a_i$ and
\begin{align*}
L_2-g(\Delta,1,h)&=L_2-\Delta\sum_{i=1}^{l}\delta(c,i-1)-(a_1-b)\delta(c,l)-h(h-c)=\sum_{i=l+2}^{l+j-1}\delta(c,i-1)(a_{i-l}-b)-h(h-c)\\
&\geq\left[\sum_{i=2}^{j-1}a_i-(j-2)b\right]\delta(c,l+j-2)-h(h-c)\geq\left[(m-j+1)b-\sum_{i=j}^{m}a_i\right]\delta(c,l+j-2)-h(h-c)\\
&\geq\sum_{i=l+j}^{l+m}\delta(c,i-1)(b-a_{i-l})-h(h-c)=g(b,b,c)-L_1-h(h-c).
\end{align*}
In both cases, $L_1+L_2\geq g(b,b,c)+g(\Delta,1,h)-h(h-c)$, and the claim follows.
\end{proof}
\paragraph{Visualization of Function $g$}
We plot the values of $g(a,b,c)$ with various values of $a$, $b$ and $c$ in Figure \ref{fig:g}. We set $a\in[100,500]$, $b=\lfloor\lambda a\rfloor$ where $\lambda\in[0.05,0.5]$ at step size $0.05$, and $c\in[10,90]$ at step size $10$. Note that $g$ is monotone.

Firstly, we can see that $g$ increases proportionally in $a$ (and $b$) under fixed $b/a$, i.e. $g(a,b,c)\in\Theta(a)$ assuming constant $b/a$ and $c$. Secondly, $g$ increases in $b/a$ at diminishing rate, and is plateaued when $b$ exceeds $\lceil a/2\rceil$ by definition. Thirdly, $g$ increases proportionally in $c^2$, i.e. $g(a,b,c)\in\Theta(c^2)$ assuming constant $a$ and $b$.

\section{Greedy Algorithms for Diverse Solutions}\label{sec:algorithm}
We describe two different greedy algorithms to obtain an approximation to the problem \eqref{eq:problem}, by incrementally building solutions. They are greedy in the sense that they select, in each step, the ``best'' choice out of a selection pool. Here, choice refers to a solution-element pair where the element is added into the solution. The differences between the two algorithms lie in how this pool is defined, and the selection criteria. In both algorithms, the pool is controlled by a parameter, which determines a trade-off between objective values and diversity.

In the following, we claim several worst-case bounds, i.e. for all settings $I$ (each including a problem instance and an algorithm parameter value) in a universe clear from the context, $p(I)\geq q(I)$ for some quantities $p$ and $q$ of the setting (e.g. optimal value, worst-case diversity, etc.) A bound is tight if there is a setting $I'$ where $p(I')=q(I')$. It is nearly tight if instead we have $p(I')=q(I')+\epsilon$ for an arbitrary small $\epsilon>0$ independent from other factors.
\subsection{Diversifying Greedy With Common Elements}
The first approach, outlined in Algorithm \ref{alg:greedy_monotone}, is a deterministic version of a heuristic for a special case of problem \eqref{eq:problem}, proposed in \cite{Neumann2021}. The idea is to first have all solutions share common elements selected by the classical greedy algorithm, so as to efficiently obtain some objective value guarantee. Then, in the second phase (starting from line \ref{greedy_monotone:initialize}), each solution is finalized with added elements that maximize $ss$, which are precisely those least represented. To be specific, in each iteration, the algorithm looks at all solution-element pairs which maintain independence, and selects a pair based on criteria, the first of which maximizes diversity (line \ref{greedy_monotone:main_selection}). This approach is simple and efficient, but prevents the common elements from contributing to diversity. Here, we formulate the algorithm to take the number of common elements as an input ($b$), which cannot exceed the rank of the matroid constraint.
\begin{algorithm}[t]
\KwIn{$f$, $S$, $b$, $r$\tcp*[f]{Assuming $b\leq\max_{z\in S}|z|$}}
\KwOut{$P\in S^r$}
$x\gets\emptyset,\kappa(\cdot)\gets\{u\in(V\setminus\cdot):(\cdot\cup\{u\})\in S\}$\;
\While{$|x|<b$ and $\kappa(x)\neq\emptyset$}{
$v\gets\argmax_{u\in\kappa(x)}f(x\cup\{u\}),x\gets x\cup\{v\}$\;
}
$P\gets\{x\}^r$\tcc*[r]{$P$ contains $r$ duplicates of $x$}
$R\gets\{(z,v):z\in P,v\in\kappa(z),f(z\cup\{v\})\geq f(z)\wedge n_v(P)<\lceil r/2\rceil\}$\;\label{greedy_monotone:initialize}
\While(\tcp*[f]{below $\argmin$ over vectors is done in left-to-right lexicographical order}){$R\neq\emptyset$}{
$(y,v)\gets\argmin_{(z,u)\in R}(n_u(P),\left|\kappa(z)\right|,f(z),f(z)-f(z\cup\{u\}))$\;\label{greedy_monotone:main_selection}
$P\gets P\setminus\{y\}\cup\{y\cup\{v\}\}$\;
Update $R$ as in Line \ref{greedy_monotone:initialize}\;
}
\caption{Greedy with common elements}
\label{alg:greedy_monotone}
\end{algorithm}

We observe that since the image of $ss$ is polynomially bounded in size, there are frequently many equivalent choices in each iteration in the second phase, motivating the use of tie-breaking rules, which are formulated as lexicographical $\argmin$ at Line \ref{greedy_monotone:main_selection}. Of note is the second rule, which prioritizes solutions with the fewest remaining choices. The idea is to minimize the shrinkage of the pool among under-represented elements (the inclusion of which incurs large marginal gains in diversity) with a simple heuristic. We show that this tie-breaking rule helps guarantee a non-trivial lower bound of $ss$ value under a general matroid, whereas it makes no difference under a uniform matroid. The other tie-breaking rules aim to improve the minimum objective value whenever possible.

The time complexity of Algorithm \ref{alg:greedy_monotone} is $O(b|V|+r(r_M-b)(|V|-b))$ in both value oracle model and independence oracle model. The algorithm may not return $r$ bases if $rr_M$ is sufficiently large relative to $|V|$. Additionally, having all solutions sharing elements can be undesirable in some applications. Note the condition $n_v(P)<\lceil r/2\rceil$ at line \ref{greedy_monotone:initialize} ensures $ss(P)$ never decreases during the second phase.

Let $\mathcal{A}(f,S,b,r)$ be the collection of possible outputs from Algorithm \ref{alg:greedy_monotone} when run with inputs $f$, $S$, $b$, $r$. We first show that in uniform constraint case, the algorithm returns a constant diversity for each input configuration.
\begin{theoremE}[][category=greedymonotoneuniform,end]\label{result:greedy_monotone_uniform}
For any monotone submodular $f$ over $2^V$, integers $K\geq1$, $r\geq2$, $b\in[0,K)$, and let $\alpha=1-e^{-b/K}$, $\forall P\in\mathcal{A}\left(f,\mathcal{U}_{V,K},b,r\right)$, $ss(P)=g(|V|-b,K-b,r)$, thus Algorithm \ref{alg:greedy_monotone} is $g(|V|-b,K-b,r)/g(|V|,K,r)$-approximate for the $\left(f,\mathcal{U}_{V,K},ss,r,\alpha\right)$-instance. Moreover, this ratio bound is tight for any $|V|\geq1$, $r\geq2$, $K\in[1,|V|]$, and $b\in[0,K)$.
\end{theoremE}
\begin{proof}
For any monotone submodular $f$ and uniform matroid $M=(V,\mathcal{I})$, the output of Algorithm \ref{alg:greedy_monotone}, $P$, contains $r$ $(1-e^{-b/K})$-approximations of $f$ over $\mathcal{I}$ \cite{Krause}. Thus, it is feasible to the $(f,ss,r,\alpha)$-instance. The value of $ss(P)$ is completely defined in the second phase.

Let $V'=V\setminus x$ be the set of remaining elements in the second phase (note that $x=cl_M(x)$ for any non-maximal $x\in\mathcal{I}$ if $M$ is uniform), $P_t$ be the multiset after $t$ steps in the second phase, which is terminated after $T>0$ steps, we have $ss(P_0)=0$. Let $n_{t,i}=|\{x\in P_t: i\in x\}|$ for $i\in V$, we now characterize $n_{T,i}$ for all $i\in V'$ in order to derive $ss(P_T)$. Since the algorithm adds one element into a solution in each step, we have $n_{t+1,i}\leq n_{t,j}$ for any $i\in V'$. Let $i_t\in V$ be the element added at step $t$, we have $\delta_t=ss(P_t)-ss(P_{t-1})=r-2n_{t-1,i_t}-1$, so $i_t\in\argmin_{i\in V'}n_{t-1,i}$ from the greedy step and the fact that $M$ is uniform. Therefore, $\max_{i\in V'}n_{t,i}=\lceil t/|V'|\rceil$ and $\min_{i\in V'}n_{t,i}=\lfloor t/|V'|\rfloor$ for all $t\in[0,T]$, implying $\delta_t=r-2\lfloor (t-1)/|V'|\rfloor-1$. Let $h=\min\{K-b,|V'|/2\}$ and $H=\lceil r/2\rceil\lceil h\rceil+\lfloor r/2\rfloor\lfloor h\rfloor$, we show $T=H$ by considering two cases:
\begin{itemize}
\item If $K-b\leq|V'|/2$, then $T=r(K-b)$ since $r$ feasible solutions cannot contain more than $rK$ elements in total, and $r-2\lfloor [r(K-b)-1]/|V'|\rfloor-1\geq r-2\lfloor r/2-1/|V'|\rfloor-1\geq0$, implying that step $r(K-b)$ does not decrease $ss$. Moreover, in this case, $H=r(K-b)=T$, proving the claim.
\item If $K-b>|V'|/2$, then $T=H<r(K-b)$ since $r-2\lfloor (H-1)/|V'|\rfloor-1\geq0$ and $r-2\lfloor H/|V'|\rfloor-1=-1$, implying that the second phase is terminated after exactly $H$ steps.
\end{itemize}
Applying Equation \eqref{eq:ss_counts}, we get $ss(P)=ss(P_T)=g(|V'|,K-b,r)=g(|V|-b,K-b,r)$, proving the claim.

The tightness follows from a simple construction. Let $s\in[1,|V|]$, $\alpha=1-e^{-b/s}$, and $f$ be any monotone submodular over $2^V$ such that $\min_{x\subseteq V:|x|=s}f(x)\geq\alpha\max_{x\subseteq V:|x|=s}f(x)$, then the multiset $P^*=\left\lbrace\left\lbrace(is+j\mod |V|)+1\right\rbrace_{j=0}^{s-1}\right\rbrace_{i=0}^{r-1}$ contains $\alpha$-approximations of $f$ over $s$-rank uniform constraint, and $ss(P^*)=g(|V|,s,r)$.
\end{proof}

Due to the monotonicity of $g$, the diversity guarantee in Theorem \ref{result:greedy_monotone_uniform} decreases with $b$. Specifically, Lemma \ref{lemma:g_sum_bound_ratio} implies this ratio bound is at least $1-b/K$, which is tight in many cases. We also observe this linear relationship frequently in our experimental results (Section \ref{sec:experiment}). This also means by setting $b$ such that $b/K$ is constant, the algorithm guarantees simultaneously constant approximation ratios in both objective values and diversity, independent of $|V|$ and $r$. Additionally, with $\alpha=1-e^{-b/K}$, we have $1-b/K=1+\ln(1-\alpha)$, giving a direct objective-diversity trade-off curve (in terms of ratios) within $\alpha\in[0,1-1/e]$.

For general matroids, we can infer the objective approximation guarantee from Algorithm \ref{alg:greedy_monotone} as a function of parameter $b$, by using an important result in \cite{Fisher1978}.
\begin{lemmaE}[][category=greedymonotone,end]
Algorithm \ref{alg:greedy_monotone} under a matroid $M=(V,\mathcal{I})$ outputs $\left[1-\left(1-1/r_M\right)^{\min\{b,k\}}\right]$-approximations of $f$ over $\mathcal{I}$ where $k=\min\left\lbrace|z|:z\in2^V\setminus\mathcal{I}\right\rbrace-1$. Additionally, these solutions are $b/(2r_M)$-approximations.
\end{lemmaE}
\begin{proof}
It suffices to show the bound for the solution $x$ obtained after the first phase, due to $f$ being monotone. The ratio $b/(2r_M)$ follows from the fact that the classical greedy guarantees $1/2$-approximation in maximizing a monotone submodular function over a matroid \cite{Fisher1978}. Let $\lambda=1-\left(1-1/r_M\right)^{\min\{b,k\}}$, $O=\argmax_{y\in\mathcal{I}}f(y)$, $X=\{y\subseteq V: x\subset y\}$, $\mathcal{J}=\mathcal{I}\setminus X$, we have that $\mathcal{J}$ is an independence system, and $x$ is a solution obtainable by the classical greedy algorithm over $\mathcal{J}$. Furthermore, by definition of $\mathcal{J}$, we have $\min\left\lbrace|z|:z\in2^V\setminus\mathcal{J}\right\rbrace=\min\{k,b\}+1$. Therefore, Theorem 2.2 in \cite{Fisher1978} implies that $f(x)\geq\lambda\max_{y\in\mathcal{J}}f(y)$. Assuming $O\setminus X\neq\emptyset$, then $\max_{y\in\mathcal{J}}\{f(y)\}=\max_{y\in\mathcal{I}}f(y)$, so the claim follows. Otherwise, let $z\in O$, then $z\supset x$, and $f(x)/f(z)\geq|x|/|z|\geq b/r_M\geq\lambda$. Indeed, let $\delta_i$ be the marginal difference from adding the $i$-th element by the classical greedy algorithm over $\mathcal{J}$, we have that for all $v\in x$ and $u\in z\setminus x$, $x\setminus\{v\}\cup\{u\}\in\mathcal{J}$, so by the greedy selection and $f$ being submodular, $f(x\cup\{v\})-f(x)\leq\delta_i$ for all $i=1,\ldots,b$. The claim follows since
\begin{align*}
f(z)&\leq f(x)+\sum_{v\in z\setminus x}\left[f(x\cup\{v\})-f(x)\right]&\hfill\text{($f$ is submodular)}\\
&\leq f(x)+|z\setminus x|\delta_b\leq f(x)+\frac{|z|-|x|}{|x|}f(x)=\frac{|z|}{|x|}f(x)&\hfill\text{($\delta_i\geq\delta_{i+1}$)}
\end{align*}
\end{proof}

\begin{theoremE}[][category=greedymonotonematroid,end]\label{result:greedy_monotone_matroid}
For any monotone $f$ over $2^V$, matroid $M=(V,\mathcal{I})$, and integers $r\geq2$, $b\in[0,r_M)$, $\forall P\in\mathcal{A}\left(f,\mathcal{U}_{V,K},b,r\right)$, $ss(P)\geq g(r_M-b-1,r_M-b-1,r)+g(m,1,r)$, where $m=|V\setminus cl_M(x)|-r_M+b+1$ and $x$ is the solution obtained in the first phase of the algorithm. Moreover, this bound is tight for any $|V|\geq1$, $r\geq2$, matroid rank $r_M\in[1,|V|]$, $b\in[0,s)$ and $m\in[1,|V|-r_M+b+1]$.
\end{theoremE}
\begin{proof}
Let $V'=V\setminus cl_M(x)$ be set of remaining elements in the second phase, we have $|V'|\geq r_M-b$. Let $P_t$ be the multiset after $t$ steps in the second phase, which is terminated after $T\leq r(r_M-b)$ steps, count values $n_{t,i}=n_i(P_t)$ for $i\in V$, and $V_t=\bigcup_{y\in P_t}V\setminus cl_M(y)\subseteq V'$ be the set of elements that can be added to a solution in $P_t$, $n_t=\min_{i\in V_t}n_{t,i}$, we see that $V_0=V'$ and for all $t\in[0,T)$, $\delta_t=ss(P_{t+1})-ss(P_t)=r-2n_t-1$ from the greedy selection. Also, $V_{i}\supseteq V_{j}$ whenever $i\leq j$, and the greedy selection implies $\min_{i\in V_t}n_{t,i}\geq\max_{i\in V_t}n_{t,i}-1$ at every step $t$, which can be shown by induction. It holds for $t=0$ as $n_{0,i}=0$ for all $i\in V'$. Assuming this holds for $t=k$, the greedy selection guarantees that the property is maintained within $V_k$ at step $k+1$, so it must hold within $V_{k+1}\subseteq V_k$ as well.

Given this invariant, we can divide the second phase into cycles: we say step $t$ is in cycle $j$ if $n_t=j$. Let $t_j$ be the first time step of cycle $j$, we see that $n_{t_j,i}=j$ for all $i\in V_{t_j}$. With the tie breaking rule, at any time $t'$, if $n_{t',i}=n'$ for all $i\in V_{t'}$, then the algorithm builds up a base in the next consecutive steps, followed by adding elements with count $n'$ until all elements in $V_{t'}$ have count $n'+1$. Let $m=|V'|-r_M+b+1$, we show, by induction, that for any $j\leq\lfloor r/j\rfloor$ there are at least $r-mj$ solutions in $P_{t_j}$ unchanged in the second phase, i.e. equal $x$. This clearly holds for $j=0$, since no solution is changed yet. If $r\geq m$, this holds for $j=1$, since in cycle 0, a base is obtained with $r_M-b$ elements in $|V'|$, leaving $m-1$ elements with count 0 to add to other solutions, resulting in at most $m$ changed solutions by the start of cycle 1. Assuming this holds for $j=k\leq\lfloor r/m\rfloor-1$, $x\in P_{t_k}$, so $V_{t_k}=V'$. This means $V_t=V'$ for all $t\in[0,t_j]$, so $t_j=j|V'|$. Now, starting from cycle $k$, the algorithm builds the next base first. Due to the tie breaking rule, the algorithm does this using a non-base solution already changed in previous cycles\footnote{If there is no such solution, then it must be that $|V'|=r_M-b$ and $m=1$. In this case, the algorithm achieves $ss$ value of $g(r_M-b,r_M-b,r)$ which is at least $g(r_M-b-1,r_M-b-1,r)+g(m,1,r)$ by Lemma \ref{lemma:g_sum_bound}.}, meaning at most $r_M-b-1$ are needed to build the base within each cycle after 0, and that $x$ remains in the set after this base is obtained. If building this base in cycle $k$ only needs $r_M-b-1-\epsilon$ steps, then this non-base contains $b+1+\epsilon$ elements at step $t_k$, so there must be at least $r-km+\epsilon$ unchanged solutions in $P_{t_k}$. Since there are $m+\epsilon$ elements with count $n_{t_k}$ afterwards, the number of unchanged solutions after cycle $k$ is at most $r-(k+1)m$. With this, for all $j\in[0,\lfloor r/m\rfloor-1]$, $t_{j+1}-t_j=|V'|$. Therefore, there are at least $l=\min\{\lfloor r/m\rfloor,\lceil r/2\rceil\}$ cycles containing $|V'|$ steps. If $l=\lceil r/2\rceil$, the algorithm terminates after $l$ cycle since subsequent steps incur negative changes to $ss$, thus achieving $ss$ value of $g(|V'|,r_M-b,r)\geq g(r_M-b-1,r_M-b-1,r)+g(m,1,r)$ where the inequality follows from Lemma \ref{lemma:g_sum_bound}. Otherwise, it must be that $m\geq2$, and we also have $ss(P_{t_{l}})=|V'|\sum_{i=1}^l\delta(r,i-1)$. If $m=2$, each step in cycle $l$ does not increase $ss$, so the algorithm achieves the same $ss$ value. Therefore, we can assume $m>2$ and $l<\lceil r/2\rceil$.

Let $c\in[0,m)$ be such that $r\equiv c\mod m$, since $t_l=l|V'|$ and $P_{t_j}$ contains at least $j$ bases, $P_{t_l}$ must contain at least $c$ unchanged solutions. If the algorithms obtains a first base in cycle $l$ in $r_M-b-1-\epsilon$ steps, then $P_{t_l}$ contains at least $c+\epsilon$ unchanged solutions. Since this base is built from a changed solution in $P_{t_l}$, there must be at least $r_M-b-1$ steps in cycle $l$. Furthermore, $t_{j}-t_{j-1}$ is non-increasing since $V_{t}$ shrinks with $t$. Also, for any $j>l$, aside from the $j$ bases built at the start of each cycle in $P_{t_j}$, the rest must contain at least in total $r-j$ elements not in $x$. This means $t_{j}\geq j(r_M-b)+r-j$ for all $j>l$, so $t_j-t_l\geq j(r_M-b)-l|V'|+r-j=(j-l)(r_M-b-1)+r-lm$. Let $b=\lceil r/2\rceil-l$ and $x_j=t_{j+l}-t_{j+l-1}$ for $j\in[1,b]$ ($x_j=0$ if the algorithm terminates before cycle $j$), then $\sum_{i=1}^{b}x_i=t_{b+l}-t_l\geq b(r_M-b-1)+r-lm$, so the conditions of Lemma \ref{lemma:greedy_monotone_matroid_lowerbound_helper} are satisfied. Therefore, applying it and the fact that each step in cycle $j$ adds $\delta(r,j)$ to the $ss$ value gives $T=t_{l+b}$ and
\begin{align*}
ss(P_T)=|V'|\sum_{i=1}^l\delta(r,i-1)+\sum_{i=l+1}^{l+b}\delta(z,i-1)x_{i-l}\geq g(r_M-b-1,r_M-b-1,r)+g(m,1,h)-h(h-r),
\end{align*}
where $h\geq lm+\sum_{i=1}^bx_i-b(r_M-b-1)\geq r$. Since $m>2$, we have then $g(m,1,h)-h(h-r)\geq g(m,1,r)$, so the claim follows.

We show tightness by construction. Let $U$ be a $(|V|-s+b-m+1)$-size subset of $V$ where $s\in[1,|V|]$, $\{U_i\}_{i=1}^{b}$ be a partitioning of $U$ where $U_i\neq\emptyset$ for all $i$, and $A=\{a_1,\ldots,a_m\}$ be a $m$-size subset of $V\setminus U$. Denoting characteristic vector of $x\subseteq V$ with $(x_i)_{i=1}^{|V|}$, let $f$ be defined over $2^V$ with $f(x)=\sum_{i\notin U}d_ix_i+\sum_{i=1}^{b}c_i\max_{j\in U_i}x_j$ where $\{d_i\}_{i\notin U}$ and $\{c_i\}_{i=1}^{b}$ are non-negative reals and $\min_ic_i\geq\max_jd_j$. Finally, let $M=(V,\mathcal{I})$ be a matroid of rank $s$ where $\mathcal{I}=\{x\subseteq V:\sum_{i=1}^mx_{a_i}\leq1\wedge\forall i=1,\ldots,b,\sum_{j\in U_i}x_j\leq1\}$. Algorithm \ref{alg:greedy_monotone}, run with inputs $f$, $ss$, $M$, $b$, $r$ achieves after the first phase a solution $x$ where $|x\cap U_i|=1$ for all $i=1,\ldots,b$. It then diversifies the solution set over $V\setminus U$ since $cl_M(x)=U$. Since each solution cannot intersect with $A$ at more than one element, the algorithm returns $P=\left\lbrace x\cup((V\setminus U)\setminus A)\cup\{a_{(\pi(i)\mod m)+1}\}\right\rbrace_{i=1}^{\lceil r/2\rceil}\cup\left\lbrace x\cup\{a_{(\pi(i)\mod m)+1}\}\right\rbrace_{i=\lceil r/2\rceil+1}^r$ for some permutation $\pi$ over $\{1,\ldots,r\}$. We have
\begin{align*}
ss(P)=ss(\{y\setminus A: y\in P\})+ss(\{y\cap A: y\in P\})=g(r_M-b-1,r_M-b-1,r)+g(m,1,r).
\end{align*}
\end{proof}
It is important to note that while the bound in Theorem \ref{result:greedy_monotone_matroid} can be small for any positive choice of $b$ if the closure of the common elements set $x$ is large, sufficiently large ones (e.g. $|cl_M(x)|/|x|>|V|/r_M$) also lower the upper bound on maximum diversity, according to Lemma \ref{lemma:optimum_bound_closure}.
\subsection{Simultaneous Greedy With Representation Limits}
The second approach, outlined in Algorithm \ref{alg:greedy_monotone2}, is inspired by the \textsc{SimultaneousGreedys} algorithm proposed in \cite{Feldman2020}, which obtains a set of disjoint solutions, in which the best one provides an approximation guarantee. Since for our problem, all solutions need to be sufficiently good, we make crucial changes to adapt the algorithm to the task. Firstly, each element can appear in multiple solutions, the maximum number of which is given as an input ($l$). This simultaneously expands the selection pool for each solution in each iteration, which helps with quality, and controls the amount of representation in the output each element enjoys, which guarantees some diversity. Secondly, a single element ($v^*$) is allowed to be included in all solutions, so a non-trivial quality guarantee is possible, as there are instances where excluding an element ensures that the solution is arbitrarily bad. Finally, the selection criteria, especially the first one, enforce building solutions evenly, so the worst one does not fall too far behind. This allows us to derive a non-trivial lower bound on the objective value of every solution in the output.

\begin{algorithm}[t]
\KwIn{$f$, $S$, $r$, $l$\tcp*[f]{Assuming $l\in\left[1,r\right]$}}
\KwOut{$P\in S^r$}
$v^*\gets\argmax_{v\in S}f(\{v\}),P\gets\{\{v^*\}\}^{r}$\;
$\kappa(\cdot)\gets\{u\in(V\setminus\cdot):(\cdot\cup\{u\})\in S\wedge n_u(P)<l\}$\;
$R\gets\left\lbrace(z,v): z\in P,v\in\kappa(z),f(z\cup\{v\})\geq f(z)\right\rbrace$\;\label{greedy_monotone2:initialize}
\While(\tcp*[f]{below $\argmin$ over vectors is done in left-to-right lexicographical order}){$R\neq\emptyset$}{
$(y,v)\gets\argmin\limits_{(z,u)\in R}(|z|,f(z)-f(z\cup\{u\}),f(z),n_u(P))$\;
$P\gets P\setminus\{y\}\cup\{y\cup\{v\}\}$\;
Update $R$ as in Line \ref{greedy_monotone2:initialize}\;
}
\caption{Greedy with representation limits}
\label{alg:greedy_monotone2}
\end{algorithm}

Compared to Algorithm \ref{alg:greedy_monotone}, this algorithm does not maximize diversity directly, but guarantees it indirectly by imposing additional constraints. Since these constraints are on elements' representation, it can be applied to the problem \eqref{eq:problem} with any diversity measure that can be formulated by elements' representation, such as entropy \cite{Neumann2021}.

The time complexity of Algorithm \ref{alg:greedy_monotone2} is $O(rr_M|V|)$ in both value oracle model and independence oracle model. Like Algorithm \ref{alg:greedy_monotone}, it may not return $r$ bases if $rr_M$ is sufficiently large and $l$ is sufficiently small. We remark that the inclusion of the initial element $v^*$ in all solutions is meant to deal with pathological instances; this might be avoided with a more complex heuristic. With a view to simplicity, we choose not to pursue this further in this work.

We observe that if $l=r$, Algorithm \ref{alg:greedy_monotone2} must return solutions obtainable by the classical greedy algorithm since if there is a $v\in V$ that cannot be added to a solution $x$ due to the new constraint, then $v\in x$. However, one can construct instances where it is guaranteed to achieve $ss$ value of 0, even when restricted to uniform matroids and linear objective functions. Therefore, we only consider cases where $l<r$. We show the extent to which diversity is guaranteed, simply from limiting elements' representations.

Similarly, we use $\mathcal{B}(f,S,r,l)$ to denote the collection of possible outputs from Algorithm \ref{alg:greedy_monotone2} when run with inputs $f$, $S$, $r$, $l$. Given a matroid $M=(V,\mathcal{I})$, for any $P\in\mathcal{B}(f,\mathcal{I},r,l)$ and $x\in P$, let $P_t$ be $P$ at iteration $t$ ($P_0=\{\{v^*\}\}^r$), $t_{x,i}$ be the iteration in which the $i$-th element is added to $x$ (counting $v^*$), $x^{(i)}$ be $x$ right after that iteration, $V_i=\left\lbrace v\in V: n_v(P_{t_{x,i}-1})\geq l\right\rbrace$, $W_i=cl_M\left(x^{(i-1)}\right)$ and $U_i=V_i\cup W_i\setminus x^{(i-1)}$. Intuitively, $U_i$ contains elements that cannot be added to $x$ at step $i$.

To establish objective value guarantees, we first extend Proposition 2.2 in \cite{Fisher1978}.
\begin{lemmaE}[][end]\label{lemma:series_pointmult_bound}
Given non-negative non-increasing $(\lambda_i)_{i=1}^T$ and non-negative $(\sigma_i)_{i=1}^T$ where $\sum_{i=1}^k\sigma_i\leq ak$ for all $k=1,\ldots,T-1$ and $\sum_{i=1}^T\sigma_i\leq aT+c$ for some $a,c\geq0$, then $\sum_{i=1}^T\sigma_i\lambda_i\leq(a+c/T)\sum_{i=1}^T\lambda_i$.
\end{lemmaE}
\begin{proof}
Let $\sigma_i'=\sigma_i$ for all $i=1,\ldots,T-1$, and $\sigma_T'=\sigma_T-c$, we have $\sum_{i=1}^k\sigma_i'\leq ak$ for all $k=1,\ldots,T$, so Proposition 2.2 in \cite{Fisher1978} implies $\sum_{i=1}^T\sigma_i'\lambda_i\leq a\sum_{i=1}^T\lambda_i$. The claim follows from $\lambda_i$ being non-increasing in $i$, which implies
\begin{align*}
\sum_{i=1}^T\sigma_i\lambda_i=\sum_{i=1}^T\sigma_i'\lambda_i+c\lambda_T\leq a\sum_{i=1}^T\lambda_i+\frac{c}{T}\sum_{i=1}^T\lambda_i=\left(a+\frac{c}{T}\right)\sum_{i=1}^T\lambda_i.
\end{align*}
\end{proof}
This gives the following helpful lower bound.
\begin{lemmaE}[][category=greedymonotone2ratio,end]\label{lemma:greedy_monotone2_ratio}
For any $Y\in(2^V)^*$, if for some $a,q\geq0$, $\sum_{v\in U_i}n_v(Y)\leq a(i-1)$ for all $i=1,\ldots,|x|$ and $\sum_{v\in U_{|x|+1}}n_v(Y)\leq a(i-1)+q$, then $\min\left\lbrace a+q/|x|+|Y|,\sum_{y\in Y}|y|\right\rbrace f(x)\geq\sum_{y\in Y}f(y)$.
\end{lemmaE}
\begin{proof}
We use similar ideas in \cite{Fisher1978}. Let $\delta_i=f\left(x^{(i)}\right)-f\left(x^{(i-1)}\right)$, we have $\delta_i\geq f\left(x^{(i-1)}\cup\{v\}\right)-f\left(x^{(i-1)}\right)$ for all $v\in U_{i+1}\setminus U_{i}$ due to greedy selection. Furthermore, since $\delta_i$ is non-increasing in $i$ due to submodularity and greedy selection for each solution in $P$, we have
\begin{align*}
\sum_{y\in Y}f(y)&\leq\sum_{y\in Y}f(y\cup x)\leq|Y|f(x)+\sum_{y\in Y}\sum_{v\in y\setminus x}[f(x\cup\{v\})-f(x)]&\hfill\text{(properties of $f$)}\\
&=|Y|f(x)+\sum_{i=2}^{|x|+1}\sum_{v\in U^i\setminus U^{i-1}}n_v(Y)[f(x\cup\{v\})-f(x)]&\hfill\text{($U^i\supseteq U^{i-1}$)}\\
&\leq|Y|f(x)+\sum_{i=1}^{|x|}\left[\sum_{v\in U_{i+1}}n_v(Y)-\sum_{v\in U_i}n_v(Y)\right]\delta_i&\hfill\text{(greedy selection)}\\
&\leq\left(a+\frac{q}{|x|}+|Y|\right)f(x),&\hfill\text{(Lemma \ref{lemma:series_pointmult_bound})}
\end{align*}
where the last inequality follows from Lemma \ref{lemma:series_pointmult_bound}. Finally, from $v^*\in x$, submodularity and monotonicity of $f$ imply
\begin{align*}
\sum_{y\in Y}f(y)\leq\sum_{y\in Y}\sum_{v\in y}f(\{v\})\leq\sum_{y\in Y}|y|\max_{v\in V}f(\{v\})=\sum_{y\in Y}|y|f(\{v^*\})\leq\sum_{y\in Y}|y|f(x).
\end{align*}
\end{proof}
We observe the following bound regarding the shrinkage of the selection pool, which when combined with the above result leads to objective value guarantees.
\begin{lemmaE}[][end]\label{lemma:greedy_monotone2_discarded_elem}
For all $i\geq1$, $\left|V_i\setminus x^{(i-1)}\right|\leq(i-1)(r-1)/l$.
\end{lemmaE}
\begin{proof}
The claim holds for $i=1$ since $V_1=\{v^*\}$. Let $z^{(i)}=x^{(i)}\setminus\{v^*\}$, we have $\emptyset=U_0=U_1\subseteq U_2\subseteq\ldots\subseteq U_{|x|+1}=V\setminus x$. Since $t_{x,\leq1}=0$, $V_i\supseteq\{v^*\}$ for $i\geq2$, and from the fact that one element is added in each iteration
\begin{align*}
\left|V_i\setminus x^{(i-1)}\right|&\leq\left\lfloor\frac{t_{x,i}-1-l\left|V_i\cap z^{(i-1)}\right|-\left|z^{(i-1)}\setminus V_i\right|}{l}\right\rfloor\\&\leq\left\lfloor\left(t_{x,i}-1-\left|z^{(i-1)}\right|\right)/l\right\rfloor&\hfill\text{($l\geq1$)}\\
&\leq\left\lfloor\left((i-1)r-1-i+2\right)/l\right\rfloor&\hfill\text{($t_{x,i}\leq(i-1)r$ due to first selection rule)}\\
&=\left\lfloor(i-1)(r-1)/l\right\rfloor\leq(i-1)(r-1)/l.
\end{align*}
\end{proof}
Regarding diversity, we show the lower bound on the $ss$ value as the algorithm progresses. 
\begin{lemmaE}[][category=greedymonotone2ss,end]\label{lemma:greedy_monotone2_ss}
For all $t\geq 0$, $ss\left(P_t\right)\geq\lfloor t/l\rfloor l(r-l)+c(r-c)$ where $c\in[0,l)$ such that $c\equiv t\mod l$.
\end{lemmaE}
\begin{proof}
We show this bound holds tightly for any $r$-size multiset $Q$ of subsets of $V$ where $\sum_{x\in Q}|x|=t$ and $\max_{u\in V}n_u(Q)\leq l$ ($v^*$ does not contribute to $ss$ value so we ignore it). This clearly holds for $t=0$. Assuming it holds tightly for all $t\in[0,k]$ for some $k\geq0$, let $c\in[0,l)$ where $c\equiv k\mod l$, and $Q$ be the $r$-size multiset such that $\sum_{x\in Q}|x|=k+1$ and $\max_{u\in V}n_u(Q)\leq l$, $v\in x$ for some $x\in Q$ and $Q'=Q\setminus\{x\}\cup\{x\setminus\{v\}\}$, we have $ss(Q)=ss(Q')+\delta(r,n_v(Q'))=ss(Q')+r-2n_v(Q')-1$, and the induction hypothesis applies to $Q'$ since $\sum_{x\in Q'}|x|=k$ and $\max_{u\in V}n_u(Q')\leq l$. If $n_v(Q')\leq c$, then
\begin{align*}
ss(Q)\geq\lfloor k/l\rfloor l(r-l)+c(r-c)+r-2c-1=\lfloor k/l\rfloor l(r-l)+(c+1)(r-c-1).
\end{align*}
If $c=l-1$, then the right hand side becomes $\lfloor(k+l)/l\rfloor l(r-l)=\lfloor(k+1)/l\rfloor l(r-l)$, otherwise we have $\lfloor(k+1)/l\rfloor=\lfloor k/l\rfloor$ (both following from definition of $c$), so the claim holds for $Q$. Note that in this case, the bound is tight if the equality holds in the induction hypothesis and $n_v(Q')=c$. Now assuming $n_v(Q')>c$, then $k\geq n_v(Q')>0$. Let $Q''=\{x\setminus\{v\}: x\in Q\}$, we have $\sum_{x\in Q''}|x|=k-n_v(Q')<k$, and using the induction hypothesis, we get
\begin{align*}
ss(Q)&=ss(Q'')+n_v(Q)[r-n_v(Q)]\geq\lfloor(k-1)/l\rfloor l(r-l)+n_v(Q)[r-n_v(Q)]+[c+l-n_v(Q')][r+n_v(Q')-h-l]\\
&\geq\lfloor(k-1)/l\rfloor l(r-l)+l(r-l)+(c+l-l+1)(r+l-1-c-l)=\lfloor(k-1+l)/l\rfloor l(r-l)+(c+1)(r-c-1)\\
&\geq\lfloor k/l\rfloor l(r-l)+(c+1)(r-c-1),
\end{align*}
where the second inequality follows from $n_v(Q')=n_v(Q)-1$ and that for any $0\leq a\leq b,c\leq d\leq e$, $a(e-a)+d(e-d)\leq b(e-b)+c(e-c)$, and the last inequality follows from $l\geq1$. Furthermore, $c<n_v(Q')=n_v(Q)-1\leq l-1$, so $\lfloor(k+1)/l\rfloor=\lfloor k/l\rfloor$, and the claim holds for $Q$, completing the induction proof.
\end{proof}
With Lemma \ref{lemma:greedy_monotone2_ratio}, \ref{lemma:greedy_monotone2_discarded_elem} and \ref{lemma:greedy_monotone2_ss}, the following objective-diversity trade-off guarantees can be inferred.
\begin{theoremE}[][category=greedymonotone2uniform,end]\label{result:greedy_monotone2_uniform}
Given monotone submodular $f$ over $2^V$, integers $r\geq2$, $l\in[1,r)$ and $K\in[1,|V|]$, then for all $P\in\mathcal{B}\left(f,\mathcal{U}_{V,K},r,l\right)$ and $k\in[1,(r-1)K/l]$,
\begin{align*}
\min\left\lbrace\frac{r-1}{l}+1,k\right\rbrace\min_{x\in P}f(x)\geq\max_{|y|\leq k}f(y)\quad\text{and}\quad ss(P)\geq l(r-l)\lfloor h/l\rfloor+c(r-c),
\end{align*}
where $h=\min\{r(K-1),l(|V|-1)\}$ and $c\in[0,l)$ such that $c\equiv h\mod l$.\footnote{The latter bound holds trivially at $l=r$.} Moreover, the former bound is nearly tight when $K\geq r+l-1$, and the latter bound is tight for any $|V|\geq1$, $r\geq2$, $K\in[1,|V|]$, and $l\in[1,r]$.
\end{theoremE}
\begin{proof}
Let $x$ be an arbitrary output solution, we see that from Lemma \ref{lemma:greedy_monotone2_discarded_elem}, for any $y\subseteq V$ where $|y|\leq k$, $|y\cap U_i|\leq(r-1)(i-1)/l$ for $i\in[1,|x|]$ since $W_i=\emptyset$, and we show that this also holds for $i=|x|+1$. If $|x|=K$, it holds since $(r-1)|x|/l=(r-1)K/l\geq k\geq|y|$ when $r\geq2$ and $l<r$. If $|x|<K$ then $W_{|x|+1}=\emptyset$, so $U_{|x|+1}=V_{|x|+1}\setminus x$. The inequality then follows directly from Lemma \ref{lemma:greedy_monotone2_discarded_elem} which asserts the upper bound on $\left|U_{|x|+1}\right|$. Combining this with Lemma \ref{lemma:greedy_monotone2_ratio} and $|y|\leq k$ yields the first inequality. For tightness, let\[f(x)=\max\{x_1,(1-\epsilon)x_{r+l}\}+\sum_{i=2}^{r+l-1}x_i-\frac{\epsilon}{lr}x_1\left(\sum_{i=2}^rx_i\right)\left(\sum_{i=r+1}^{r+l-1}x_i\right),\]observe $f$ is monotone submodular for any $\epsilon\in[0,1]$. Assuming w.l.o.g. that $\epsilon>0$, $v^*=1$ and that the last tie-breaking prioritizes smallest labels, there must be one solution $x'$ returned by Algorithm \ref{alg:greedy_monotone2} whose second element is in $\{r+1,\ldots,r+l-1\}$ while the rest's are contained in $\{2,\ldots,r\}$ due to the fourth tie-breaking rule. After adding the $l$-th element to each solution, $x'^{(l)}=\{1,r+1,\ldots,r+l-1\}$, meaning any element available to add to $x'$ at this point induces a marginal gain less than $1$ with $\epsilon>0$. Since other solutions currently do not contain elements in $\{r+1,\ldots,r+l\}$, and elements in $\{2,\ldots,r\}$ can still be added to them, inducing marginal gains $1$, the algorithm prioritizes them due to the second tie-breaking rule\footnote{Note that swapping the second and third rules makes no difference at this point as all solutions have the same objective value.}. When finally adding the $l+1$-th element to $x'$, $n_v(P)=l$ for all $v\in\{2,\ldots,r\}$, so $x'$ enjoys no further marginal gain, resulting in $f(x')=l$. We see that for $K\geq r+l-1$, $y=\{2,\ldots,r+l\}$ is feasible under $K$-rank uniform constraint and $f(y)=r+l-1-\epsilon=[(r-1)/l+1]f(x')-\epsilon$.

The second inequality follows from Lemma \ref{lemma:greedy_monotone2_ss} and that the Algorithm \ref{alg:greedy_monotone2} runs for exactly $h=\min\{r(K-1),l(|V|-1)\}$ iterations. To show tightness, let $f(x)=\sum_{i=1}^{|V|}c_ix_i$ where $(x_i)_{i=1}^{|V|}$ is the characteristic vector of $x\subseteq V$ and $(c_i)_{i=1}^{|V|}$ is a decreasing non-negative real sequence, Algorithm \ref{alg:greedy_monotone2} run on this instance under $K$-rank uniform constraint returns $P$ such that $n_1(P)=r$, $n_v(P)=l$ for $v=2,\ldots,\lfloor h/l\rfloor$, $n_{\lfloor h/l\rfloor+1}(P)=c$ where $c\in[0,l)$ such that $c\equiv h\mod l$, and $n_v(P)=0$ for $v\geq\lfloor h/l\rfloor+2$. This means $ss(P)=\lfloor h/l\rfloor l(r-l)+c(r-c)$.
\end{proof}

\begin{theoremE}[][category=greedymonotone2matroid,end]\label{result:greedy_monotone2_matroid}
Given monotone submodular $f$ over $2^V$, matroid $M=(V,\mathcal{I})$, integers $r\geq2$ and $l\in[1,r)$, then for all $P\in\mathcal{B}\left(f,\mathcal{I},r,l\right)$
\begin{align*}
\min\left\lbrace\frac{r-1}{l}+2,r_M\right\rbrace\min_{x\in P}f(x)\geq\max_{y\in\mathcal{I}}f(y)\quad\text{and}\quad ss(P)\geq l(r-l)(r_M-1).
\end{align*}
Moreover, the former bound is nearly tight, and the latter bound is tight for any $|V|\geq1$, $r\geq2$, matroid rank $r_M\in[1,|V|]$, and $l\in[1,r]$.
\end{theoremE}
\begin{proof}
We have for $i=1,\ldots,|x|+1$, $r_M(W_i)=r_M\left(x^{(i-1)}\right)=i-1$ by definition, and 
\begin{align*}
r_M\left(U_i\right)&\leq r_M\left(V_i\setminus x^{(i-1)}\right)+r_M\left(W_i\setminus x^{(i-1)}\right)-r_M\left(V_i\cap W_i\setminus x^{(i-1)}\right)&\hfill\text{($r_M$ is submodular)}\\
&\leq\left|V_i\setminus x^{(i-1)}\right|+i-1&\hfill\text{($r_M$ is monotone)}\\
&\leq[\left(r-1\right)/l+1](i-1),&\hfill\text{(Lemma \ref{lemma:greedy_monotone2_discarded_elem})}
\end{align*}
where the last inequality follows from Lemma \ref{lemma:greedy_monotone2_discarded_elem}. This means for any $y\in\mathcal{I}$, $\left|y\cap U_i\right|\leq[\left(r-1\right)/l+1](i-1)$ by Lemma \ref{lemma:matroid_freedom_limit}. Applying Lemma \ref{lemma:greedy_monotone2_ratio} and $|y|\leq r_M$ yields the first claim. For tightness, let\[f(x)=\max\{x_1,(1-\epsilon)x_{l+1}\}+\sum_{i=2}^l\max\{x_i,x_{i+l}\}+\left(1-\frac{\epsilon x_1}{lr}\right)\sum_{i=2l+1}^{3l}x_i+\sum_{i=3l+1}^{3l+r-1}x_i-\frac{\epsilon x_1}{lr}\left(\sum_{i=2}^{l}\max\{x_i,x_{i+l}\}\right)\left(\sum_{i=3l+1}^{3l+r-1}x_i\right),\]observe $f$ is monotone submodular for any $\epsilon\in[0,1]$. We construct the matroid $M=(V,\mathcal{I})$ where $\mathcal{I}=\{x\subseteq V:\sum_{i=1}^l(x_i+x_{i+2l})\leq l\}$. Assuming w.l.o.g. that $\epsilon>0$, $v^*=1$, it can be the case that there is one solution $x'$ returned by Algorithm \ref{alg:greedy_monotone2} under $M$ whose second element is in $\{2,\ldots,l\}$ while the rest's are contained in $\{3l+1,\ldots,3l+r-1\}$ due to the fourth tie-breaking rule. After adding the $l$-th element to each solution, we can assume that $x'^{(l)}=\{1,\ldots,l\}$, meaning any element available to add to $x'$ at this point induces a marginal gain less than $1$. Since other solutions currently do not contain elements in $\{2,\ldots,3l\}$, and elements in $\{3l+1,\ldots,3l+r-1\}$ can still be added to them, inducing marginal gains $1$, the algorithm prioritizes them due to the second tie-breaking rule. Note that elements in $\{l+1,\ldots,3l\}$ are ignored in the next $r-1$ iterations since they induce marginal gains less than $1$ with $\epsilon>0$. When finally adding the $l+1$-th element to $x'$, $n_v(P)=l$ for all $v\in\{3l+1,\ldots,3l+r-1\}$, and elements in $\{2l+1,\ldots,3l\}$ cannot be considered, so $x'$ enjoys no further marginal gain, resulting in $f(x')=l$. We have $y=\{l+1,\ldots,3l+r-1\}\in\mathcal{I}$ and $f(y)=r+2l-1-\epsilon=[(r-1)/l+2]f(x')-\epsilon$.

For the second claim, we see that since the algorithm runs for at least $l(r_M-1)$ iterations, we have $ss(P)\geq l(r-l)(r_M-1)$ from Lemma \ref{lemma:greedy_monotone2_ss} and that the $ss$ value cannot decrease below $l(r-l)(r_M-1)$ during iterations after $l(r_M-1)$. To show tightness, let $f(x)=\sum_{i=1}^{|V|}c_ix_i$ where $(x_i)_{i=1}^{|V|}$ is the characteristic vector of $x\subseteq V$ and $(c_i)_{i=1}^{|V|}$ is a decreasing non-negative real sequence, $M=(V,\mathcal{I})$ be a $s$-rank matroid to be constructed where $s\in[1,|V|]$, and $P$ be an output of Algorithm \ref{alg:greedy_monotone2} when run with inputs $f$, $\mathcal{I}$, $r$, $l$. For matroid $M$, we specify $\mathcal{I}=\left\lbrace x\subseteq V: x_1+\sum_{i=s+1}^{|V|}\leq1\right\rbrace$, then since for all $y\in P$, $1\in y$, we have $y\cap\{s+1,\ldots,|V|\}=\emptyset$, thus $ss(P)=l(r-l)(s-1)$.
\end{proof}

We remark that the tightness cases in the proof of Theorem \ref{result:greedy_monotone2_matroid} prevent Algorithm \ref{alg:greedy_monotone2} from exercising the last tie-breaking rule, which is the component that lets it improve diversity beyond the lower bound. We suspect that this bound might be overly pessimistic for instances where the image under $f$ of the feasible set is small.

The result suggests that for uniform constraints, setting $l=\max\{\lfloor r(r_M-1)/(|V|-1)\rfloor,1\}$ leads to Algorithm \ref{alg:greedy_monotone2} guaranteeing $(1-1/|V|)(1-O(1/r_M))$ approximation ratio in diversity, whereas $l=\lfloor r/2\rfloor$ guarantees $(r_M-1)/|V|$ approximation ratio for pathological matroid constraint. Additionally, if $l/r$ is constant, then every output solution guarantees a constant approximation ratio in objective value.

Above results only consider extreme values (e.g. optimal $f$ value). On the other hand, by comparing the algorithm's output against an arbitrary solution set, a more nuanced picture emerges which suggests the algorithm can exploit a certain feature in the global structure of $f$ to lessen compromise on diversity (i.e. by lowering parameter $l$) while maintaining objective guarantees.
\begin{theoremE}[][category=greedymonotone2uniformP,end]\label{result:greedy_monotone2_uniform_P}
Given monotone submodular $f$ over $2^V$, integers $r\geq2$, $l\in[1,r)$, $K\in[1,|V|]$ and $Y\in(2^V)^*$ such that $m=\max_{v\in V}n_v(Y)$, then for all $P\in\mathcal{B}\left(f,\mathcal{U}_{V,K},r,l\right)$
\begin{align*}
\min\left\lbrace\frac{m(r-1)h}{l}+|Y|,\sum_{y\in Y}|y|\right\rbrace\min_{x\in P}f(x)\geq\sum_{y\in Y}f(y),
\end{align*}
where $h=\max\left\lbrace l\sum_{y\in Y}|y|/[Km(r-1)],1\right\rbrace$. This bound is nearly tight for all $r\geq2$, $l\in[1,r)$, size of $Y$ and $m\in[1,|Y|]$.
\end{theoremE}
\begin{proof}
Let $x$ be any solution in $P$, since $n_v(Y)\leq m$ for all $v\in V$, $\sum_{v\in U_i}n_v(Y)\leq m|U_i|$. For all $i=1,\ldots,|x|$, $W_i=\emptyset$ so by Lemma \ref{lemma:greedy_monotone2_discarded_elem}, $\sum_{v\in U_i}n_v(Y)\leq m(r-1)(i-1)/l$. If $|x|<K$, the same inequality holds for $i=|x|+1$, otherwise the inequality holds for $i=|x|+1$ iff $h=1$, so thus far the claim holds according to Lemma \ref{lemma:greedy_monotone2_ratio}. Assuming $h>0$, then $h'=\sum_{y\in Y}|y|-Km(r-1)/l>0$, Lemma \ref{lemma:greedy_monotone2_ratio} implies $[m(r-1)/l+h'/K+|Y|]f(x)\geq\sum_{y\in Y}f(y)$. Reformulating $h'$ in $h$ gives the desired expression.

For tightness, we fix $k\geq m\geq1$, let $w=\lceil k/m\rceil$ and\[f(x)=\max\left\lbrace x_1,\left(1-\frac{\epsilon}{k}\right)z_{r+l}\right\rbrace+\sum_{i=2}^{r}x_i+\sum_{i=r+1}^{r+l-1}z_i-\frac{\epsilon x_1}{klr}\left(\sum_{i=2}^rx_i\right)\left(\sum_{i=r+1}^{r+l-1}z_i\right),\]where $z_i=\max\{x_{jl+i}\}_{j=0}^{w-1}$, $f$ is monotone submodular for any $\epsilon\in[0,k]$. Assuming w.l.o.g. that $\epsilon>0$, $v^*=1$ and that the last tie-breaker prioritizes smallest labels, there must be one solution $x'$ returned by Algorithm \ref{alg:greedy_monotone2} whose second element is in $\{r+1,\ldots,r+l-1\}$ while the rest's are contained in $\{2,\ldots,r\}$ due to the fourth tie-breaking rule. After adding the $l$-th element to each solution, $x'^{(l)}=\{1,r+1,\ldots,r+l-1\}$, meaning any element available to add to $x'$ at this point induces a marginal gain less than $1$ with $\epsilon>0$. Since other solutions currently do not contain elements in $\{r+1,\ldots,r+wl\}$, and elements in $\{2,\ldots,r\}$ can still be added to them, inducing marginal gains $1$, the algorithm prioritizes them due to the second tie-breaking rule. When finally adding the $l+1$-th element to $x'$, $n_v(P)=l$ for all $v\in\{2,\ldots,r\}$, so $x'$ enjoys no further marginal gain, resulting in $f(x')=l$. For $i=0,\ldots,k-1$, let $y_i=\{(i\mod w)l+j\}_{j=r+1}^{r+l}$, and let $Y=\left\lbrace y_i\cup\{2,\ldots,r\}\right\rbrace_{i=0}^{m-1}\cup\left\lbrace y_i\right\rbrace_{i=m}^{k-1}$, we have $m=\max_{v\in V}n_v(Y)$ and $\sum_{y\in Y}f(y)=m(r-1)+k(l-\epsilon/k)=[m(r-1)/l+k]f(x')-\epsilon=\sum_{y\in Y}|y|-\epsilon$. Thus the bound is nearly tight for $K\geq kl^2/[m(r-1)]+l$, in which case $|x'|\geq l$ when the algorithm is run under $K$-rank uniform constraint.
\end{proof}

\begin{corollaryE}[][category=greedymonotone2uniformPcol,end]
If there is $Y\in(2^V)^k$ for some $k\geq1$ where $\max_{v\in V}n_v(Y)<l\sum_{y\in Y}|y|/[K(r-1)]$ and $\sum_{y\in Y}f(y)/k\geq\alpha\max_{|y|\leq K}f(y)$, then Algorithm \ref{alg:greedy_monotone2} under $K$-rank uniform constraint returns $\alpha/2$-approximations with parameter $l$. If there is a set of $k$ disjoint $\alpha$-approximations, then Algorithm \ref{alg:greedy_monotone2} returns $\alpha/2$-approximations at any $l\in[(r-1)/k,r)$.
\end{corollaryE}

\begin{theoremE}[][category=greedymonotone2matroidP,end]\label{result:greedy_monotone2_matroid_P}
Given monotone submodular $f$ over $2^V$, integers $r\geq2$, $l\in[1,r)$, matroid $M=(V,\mathcal{I})$ and $Y\in\mathcal{I}^*$ such that $m=\max_{v\in V}n_v(Y)$, then for all $P\in\mathcal{B}\left(f,\mathcal{I},r,l\right)$
\begin{align*}
\min\left\lbrace\frac{m(r-1)}{l}+2|Y|,\sum_{y\in Y}|y|\right\rbrace\min_{x\in P}f(x)\geq\sum_{y\in Y}f(y).
\end{align*}
This bound is nearly tight for all $r\geq2$, $l\in[1,r)$, size of $Y$ and $m\in[1,|Y|]$.
\end{theoremE}
\begin{proof}
For any $x\in P$, we see from Lemma \ref{lemma:matroid_freedom_limit} that for all $y\in Y$, $|y\cap W_i|\leq i-1$ for $i=1,\ldots,|x|+1$. Therefore, $\sum_{v\in W_i}n(Y)\leq|Y|(i-1)$ for $i=1,\ldots,|x|+1$, meaning $\sum_{v\in U_i}n(Y)\leq[m(r-1)/l+|Y|](i-1)$. This yields the claim as implied by Lemma \ref{lemma:greedy_monotone2_ratio}.

For tightness, we fix $k\geq1$ and $m\in[1,k]$, let $w=\lceil k/m\rceil$ and
\begin{align*}
f(x)&=\max\left\lbrace x_1,\left(1-\frac{\epsilon}{k}\right)z_{l+1}\right\rbrace+\sum_{i=2}^l\max\{x_i,z_{i+l}\}+\left(1-\frac{\epsilon x_1}{klr}\right)\sum_{i=(w+1)l+1}^{(w+2)l}z_i\\&+\sum_{i=(w+2)l+1}^{(w+2)l+r-1}x_i-\frac{\epsilon x_1}{klr}\left(\sum_{i=2}^{l}\max\{x_i,z_{i+l}\}\right)\left(\sum_{i=(2w+1)l+1}^{(2w+1)l+r-1}x_i\right),
\end{align*}
where $z_i=\max\{x_{jl+i}\}_{j=0}^{w-1}$, $f$ is monotone submodular for any $\epsilon\in[0,k]$. We construct the matroid $M=(V,\mathcal{I})$ where $\mathcal{I}=\{x\subseteq V:\sum_{i=1}^lx_i+\sum_{i=1}^{wl}x_{(w+1)l+i}\leq l\}$. Assuming w.l.o.g. that $\epsilon>0$, $v^*=1$, it can be the case that there is one solution $x'$ returned by Algorithm \ref{alg:greedy_monotone2} under $M$ whose second element is in $\{2,\ldots,l\}$ while the rest's are contained in $\{(2w+1)l+1,\ldots,(2w+1)l+r-1\}$ due to the fourth tie-breaking rule. After adding the $l$-th element to each solution, we can assume that $x'^{(l)}=\{1,\ldots,l\}$, meaning any element available to add to $x'$ at this point induces a marginal gain less than $1$. Since other solutions currently do not contain elements in $\{2,\ldots,(2w+1)l\}$, and elements in $\{(2w+1)l+1,\ldots,(2w+1)l+r-1\}$ can still be added to them, inducing marginal gains $1$, the algorithm prioritizes them due to the second tie-breaking rule. Note that elements in $\{l+1,\ldots,(2w+1)l\}$ are ignored in the next $r-1$ iterations since they induce marginal gains less than $1$ with $\epsilon>0$. When finally adding the $l+1$-th element to $x'$, $n_v(P)=l$ for all $v\in\{(2w+1)l+1,\ldots,(2w+1)l+r-1\}$, and elements in $\{(w+1)l+1,\ldots,(2w+1)l\}$ cannot be considered, so $x'$ enjoys no further marginal gain, resulting in $f(x')=l$. For $i=0,\ldots,k-1$, let $y_i=\{(i\mod w)l+j,[(i\mod w)+w]l+j\}_{j=l+1}^{2l}$, we have $y_i\cup\{(2w+1)l+1,\ldots,(2w+1)l+r-1\}\in\mathcal{I}$. Let $Y=\left\lbrace y_i\cup\{(2w+1)l+1,\ldots,(2w+1)l+r-1\}\right\rbrace_{i=0}^{m-1}\cup\left\lbrace y_i\right\rbrace_{i=m}^{k-1}$, we have $m=\max_{v\in V}n_v(Y)$, and $\sum_{y\in Y}f(y)=m(r-1)+k(2l-\epsilon/k)=[m(r-1)/l+2k]f(x')-\epsilon=\sum_{y\in Y}|y|-\epsilon$.
\end{proof}

\begin{corollaryE}[][category=greedymonotone2matroidPcol,end]
Given a matroid $M=(V,\mathcal{I})$, if there is $Y\in\mathcal{I}^k$ for some $k\geq1$ where $\max_{v\in V}n_v(Y)\leq lk/(r-1)$ and $\sum_{y\in Y}f(y)/k\geq\alpha\max_{y\in\mathcal{I}}f(y)$, then Algorithm \ref{alg:greedy_monotone2} under matroid constraint $M$ returns $\alpha/3$-approximations with parameter $l$. If there is a set of $k$ disjoint $\alpha$-approximations, then Algorithm \ref{alg:greedy_monotone2} returns $\alpha/3$-approximations at any $l\in[(r-1)/k,r)$ and $\alpha/(2+1/k)$-approximations at $l=r-1$.
\end{corollaryE}
Going further, these bounds can be strictly improved when the number of disjoint optimal solutions exceeds certain thresholds. In particular, we show that in such cases, Algorithm \ref{alg:greedy_monotone2} guarantees objective values identical to those from the classical greedy when maximizing monotone submodular functions under the same constraints \cite{Nemhauser1978,Fisher1978}. For a function $f$ and a solution set $S$ let $D(f,S,\alpha)$ be the largest number of disjoint non-empty $\alpha$-approximations of $f$ over $S$, and for a solution $x$, let $i_x$ be its size before it stops being improved by the algorithm.
\begin{theoremE}[][category=greedymonotone2disjoint,end]\label{result:greedy_monotone2_disjoint}
Given monotone submodular $f$ over $2^V$, integers $r\geq2$, $l\in[1,r)$, and matroid $M=(V,\mathcal{I})$, then for all $P\in\mathcal{B}\left(f,\mathcal{I},r,l\right)$, given $x\in P$ where $|x|>1$ and $D(f,\mathcal{I},\alpha)>\lfloor\eta(r-1)/l\rfloor$ for some $\alpha$, then
\begin{itemize}
\item $f(x)\geq\alpha\left[1-\left(1-1/|x|\right)^{|x|}\right]\max_{y\in\mathcal{I}}f(y)$ if $M$ is uniform and $\eta=|x|-1$,
\item $f(x)\geq\alpha\max_{y\in\mathcal{I}}f(y)/2$ if $M$ is non-uniform and $\eta=i_x$.
\end{itemize}
If $D(f,\mathcal{I},\alpha)=\lfloor\eta(r-1)/l\rfloor$, the bound does not necessarily hold in either case.
\end{theoremE}
\begin{proof}
Let $Y$ be the set of such $\alpha$-approximations. In uniform case Lemma \ref{lemma:greedy_monotone2_discarded_elem} and the pigeonhole principle imply that there must be a solution $y\in Y$ where $y\cap V_i\setminus x^{(i-1)}=\emptyset$ for all $i=1,\ldots,|x|$. Since $M$ is uniform, $W_i\setminus x^{(i-1)}=\emptyset$ for all $i=1,\ldots,|x|$, so no element in $y$ is discarded when building $x$, meaning every greedy marginal improvement to $x$ is at least as good as improvement by adding any element in $y$ at corresponding steps. This means classical greedy improvement arguments hold, giving
\begin{align*}
f\left(y\cup x^{(i)}\right)-f\left(x^{(i)}\right)\leq\sum_{v\in y}\left[f\left(x^{(i)}\cup\{v\}\right)-f\left(x^{(i)}\right)\right]\leq\frac{1}{|y|}\left[f\left(x^{(i+1)}\right)-f\left(x^{(i)}\right)\right].
\end{align*}
From this, we see the distance between $f(y)$ and $f$ value of partial $x$ is multiplied by at most $1-1/|y|$ in each step, leading to the final bound $f(x)\geq\left[1-(1-1/|y|)^{|x|}\right]f(y)$. If $|y|>|x|$ and there is no $1$-size solution in $Y$, $|V|\geq(|x|-1)(r+l-1)/l+\lfloor(|x|-1)(r-1)/l\rfloor+1$. As $|x|>1$, this means $V$ has enough elements for the algorithm to make $x$ $(|x|+1)$-size, so either $|y|\leq|x|$ or $Y$ contains a $1$-size solution. The former case implies the ratio is lower bounded by $1-(1-1/|x|)^{|x|}$, while the latter gives $f(x)\geq\alpha\max_{z\in\mathcal{I}}f(z)$, both satisfying the first inequality.

In non-uniform case, we have $y\cap V_{i_x+1}=\emptyset$, and $\left|y\cap W_i\right|\leq i-1$ for all $i=1,\ldots,|x|+1$ by Lemma \ref{lemma:matroid_freedom_limit}. So by Lemma \ref{lemma:greedy_monotone2_ratio}, $f(x)=f\left(x^{(i_x)}\right)\geq f(y)/2$, yielding the second inequality.

We show the negative claim by construction. For $K$-rank uniform constraint, let $h=\lfloor(K-1)(r+l-1)/l\rfloor$, $\lambda=1-1/K$,\[f^*(x)=1-\lambda^{-s_x}+\frac{\lambda^{-s_x}}{K}\sum_{i=0}^{h-K}x_{K+i}+\left(\frac{\lambda^{-s_x}}{K}-\epsilon x_1\right)\sum_{i=1}^{K-1}z_{h+i}-\epsilon\sum_{i=2}^{K-1}x_i\sum_{i=0}^{h-K}x_{K+i},\]where $z_i=\max\{x_{i+j(K-1)}\}_{j=0}^{h-K}$ and $s_x=\sum_{i=1}^{K-1}x_i$, and $f(x)=\min\{f^*(x),1\}$, $f$ is monotone submodular with sufficiently small positive $\epsilon$ ($\epsilon<\lambda^K/(hK^2)$ suffices). Assuming $K>2$, $f(\{1\})=1/K=\max_{v}f(\{v\})$, so we can assume $v^*=1$. For the second element, we see that $f(\{1,i\})=1-\lambda^2$ for $i=2,\ldots,h$ and $f(\{1,i\})\leq1-\lambda^2-\epsilon$ for $i>h$, so we can assume Algorithm \ref{alg:greedy_monotone2} adds elements in $\{K,\ldots,h\}$ as second elements to $r-1$ solutions, while one solution has its second element in $\{2,\ldots,K-1\}$. Let $z$ be this solution, the next $K-3$ elements added to $z$ must be in $\{2,\ldots,K-1\}$ while the other solutions have subsequent elements in $\{K,\ldots,h\}$ with $\epsilon>0$ due to the second tie-breaking rule. When adding the last element to $z$, elements in $\{K,\ldots,h\}$ cannot be considered since they must first be added to other solutions due to the second tie-breaking rule until there are $l$ solutions containing each of them, and there are only enough of them to finalize up to $r-1$ solutions. Thus, $f(z)=1-\lambda^K-\epsilon$ and $|z|=K$. For $i=0,\ldots,h-K$, let $y_i=\{K+i\}\cup\{h+i(K-1)+j\}_{j=1}^{K-1}$, $y_i$ are disjoint and $f(y_i)=1>f(z)/\left(1-\lambda^K\right)$, so the bound does not hold. Note that we cannot get more than $h-K+1$ disjoint solutions to this instance with objective values $1$.

For non-uniform case, let $K>2$, $h=\lfloor K(r+l-1)/l\rfloor$, $q=h+K\lfloor K(r-1)/l\rfloor$,\[f(x)=\max\{x_1,(1-\epsilon)z_{q+1,K-1}\}+\sum_{i=2}^{K}\max\{x_i,z_{q+i}\}+\sum_{i=K}^{h}x_i+\sum_{i=1}^Kz_{h+i}-\frac{\epsilon x_1}{2hK}\left(\sum_{i=2}^{K}\max\{x_i,z_{q+i}\}+\sum_{i=1}^Kz_{h+i}\right)\sum_{i=K}^{h}x_i,\]where $z_{i}=\max\{x_{i+j(K-1)}\}_{j=0}^{h-K}$, $f$ is monotone submodular with sufficiently small positive $\epsilon$ ($\epsilon<1/h$ suffices). Finally, let $\mathcal{I}=\left\lbrace x\subseteq V:\sum_{i=1}^{K}x_i+\sum_{i=h+1}^{q}x_i\leq K\right\rbrace$. Again, we can assume $v^*=1$, and $r-1$ solutions have their second elements in $\{K+1,\ldots,h\}$ while one solution $z$ has it in $\{2,\ldots,K\}$; the same goes for their next $K-3$ elements due to the second tie-breaking rule and $\epsilon>0$. We see that the algorithm can add an element in $\{2,\ldots,K\}$ to $z^{(K-1)}$, followed by adding elements in $\{K+1,\ldots,h\}$ to other solutions until each is contained in $l$ solutions. At this point, the matroid constraint prevent elements in $\{K,\ldots,q\}$ from being added to $z$, so no further improvement occurs, leading to $f(z)=K=i_z$. For $i=0,\ldots,h-K-1$, let $y_i=\{K+i+1\}\cup\{h+iK+j,q+iK+j\}_{j=1}^{K}$, $y_i$ are disjoint, $y_i\in\mathcal{I}$ and $f(y_i)=2K+1-\epsilon>2f(z)$ given that $\epsilon<1$, so the bound does not hold. Note that we cannot get more than $h-K$ disjoint solutions to this instance with objective values greater than $2K$.
\end{proof}
We include a simple observation relating maximum diversity and the number of disjoint approximations.
\begin{observation}
Let $k=D(f,S,\alpha)$ for some function $f$, solution set $S\subseteq2^V$, threshold $\alpha$, then the maximum $ss$ value to a $(f,S,r,\alpha)$-instance is at most $g(h,s,r)$ where $h=\min\{r(s-1)+k,|V|\}$ and $s=\max_{x\in S}|x|$.
\end{observation}

\section{Experimental Investigation}\label{sec:experiment}

To observe how these algorithms perform on concrete instances, we experiment with the maximum vertex coverage problem: given a graph $G=(V,E)$, find a set $x\in\mathcal{I}$ for some matroid $M=(V,\mathcal{I})$ that maximizes $|x\cup\{v\in V:\exists u\in x,\{u,v\}\in E\}|$, which is monotone submodular. For the benchmark instance, we use the complement of frb30-15-1, frb30-15-2, frb35-17-1, frb40-19-1 from the standard benchmark suite BHOSLIB created using the Model RB \cite{BHOSLIB}, containing 450, 450, 595, 760 vertices respectively, and 17827, 17874, 27856, 41314 edges respectively; these are available at \cite{nr}.\ignore{ and licensed by CC BY-SA. The experiments were run on a hexa-core 2.2 GHz Intel i7 CPU with 16 GB of RAM.}

We use four matroid constraints in the experiments, including 2 uniform matroids and 2 partition matroids. As mentioned, partition matroids admit a independence collections of the form $\mathcal{I}=\{x\subseteq V:\forall i=1,\ldots,k,|x\cap V_i|\leq b_i\}$ for some partitioning $\{V_i\}_{i=1}^k$ of $V$ and integers $\{b_i\}_{i=1}^k$. These are useful in modeling group-based budget constraints \cite{Cornuejols1977,Nemhauser1978,Chekuri2004,Chekuri2005,Fleischer2006}. For uniform matroids, we set the ranks to $\{10,15\}$, and denote them with U10 and U15, respectively (the numbers represent the ranks). For partition matroids, we group consecutive vertices sorted by degrees into 10 partitions, i.e. $V_i$ contains from $|V|(i-1)/10+1$-th to $|V|i/10$-th smallest degree vertices. This is to force the solutions to include a limited number of high-degree vertices, creating scenarios where the greedy algorithm would obtain very different solutions from the ones it would return under uniform constraints. In case 10 does not divide $|V|$, we set $|V_i|=\lfloor|V|/10\rfloor$ for $i=2,\ldots,10$. For the first partition matroid (denoted P10), we set $b_i$ to $1$ for all $i$, while for the second (denoted P15), we assign $6$ to $b_1$ and $1$ to the rest.

For each of the 16 instances, we run with $r\in\{20,100\}$, Algorithm \ref{alg:greedy_monotone} with all parameter values $b\in[0,r_M]$, and Algorithm \ref{alg:greedy_monotone2} with all parameter values $l\in[1,r]$. For both algorithms, the last tie-breaking is done by selecting the first choice (in increasing order of vertex labels). Therefore, there is no randomization, so each algorithm is run once on each instance with each parameter value.

To contextualize the results, we obtain a best known coverage for each instance using the built-in integer linear programming solver in MATLAB. Furthermore, the upper bounds on $ss$ values are given by $\sum_{i}g(|V_i|,b_i,r)$ since $ss$ can be decomposed by disjoint subsets (in case of uniform matroid, $V_i\gets V$ and $b_i\gets r_M$). Note that this bound applies to all threshold ratio $\alpha\in[0,1]$, the actual optimal value might very well be much smaller, especially for $\alpha$ close to 1. We choose not to normalize our results against actual optimal values because \begin{enumerate*}[label=\itshape\alph*\upshape)]
\item solving exactly the problem \eqref{eq:problem} is prohibitively costly, and
\item this is exacerbated by the large number of $\alpha$ values for each of which an optimal value needs to be obtained (i.e. the number of distinct minimum objective values from the algorithms on each instance).
\end{enumerate*}

\begin{figure}[t!]
\centering
\includegraphics[width=.85\linewidth]{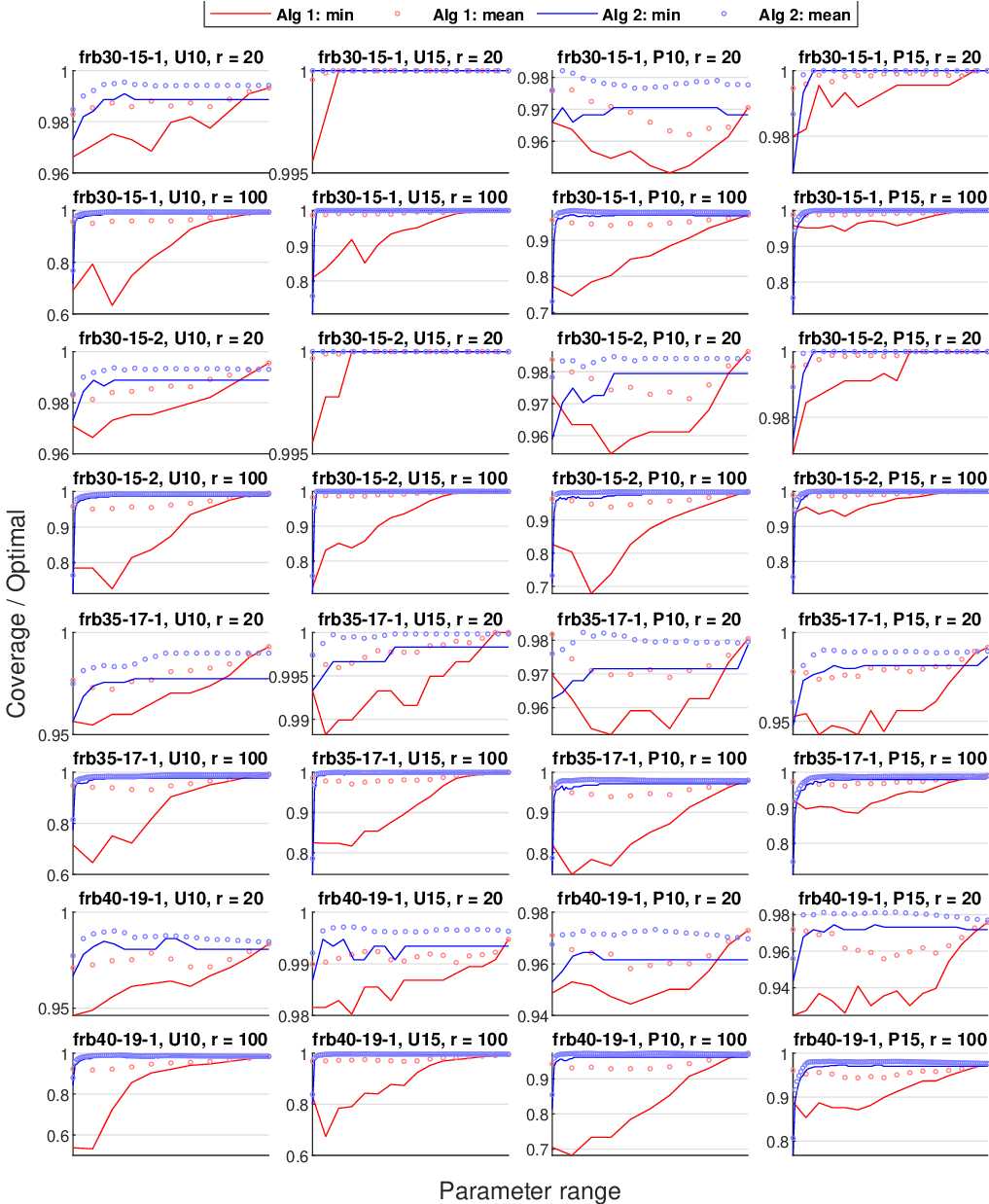}\vspace{-5pt}
\caption{Minimum and mean objective values in the outputs of the algorithms run with all parameter values in respective ranges: $[0,r_M]$ for Algorithm \ref{alg:greedy_monotone}, $[1,r]$ for Algorithm \ref{alg:greedy_monotone2}. Values are normalized against respective known optima.}\label{fig:quality}
\end{figure}

\begin{figure}[t!]
\centering
\includegraphics[width=.9\linewidth]{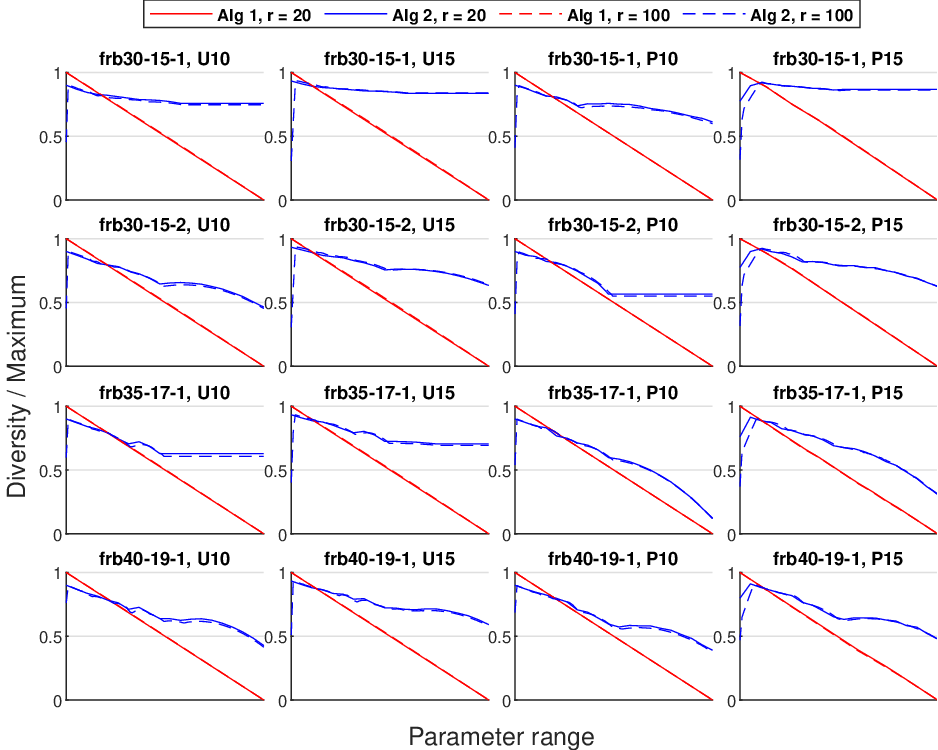}\vspace{-5pt}
\caption{Diversity as $ss$ values of the outputs of the algorithms run with all parameter values in respective ranges: $[0,r_M]$ for Algorithm \ref{alg:greedy_monotone}, $[1,r]$ for Algorithm \ref{alg:greedy_monotone2}. Values are normalized against respective known upper bounds.}\label{fig:diversity}
\end{figure}

The results are shown in Figure \ref{fig:quality} and \ref{fig:diversity}, which visualize, for each graph-constraint-parameter combination, the mean and minimum objective values in the output, and the $ss$ value. We see that the objective values in the outputs are high (within 5\% gap of the optimal) for $r=20$, and predictably degrade for $r=100$ (about 30\%), although the mean values stay within 10\% in most settings. Notably, Algorithm \ref{alg:greedy_monotone2} produces higher minimum objective values than Algorithm \ref{alg:greedy_monotone} does in most settings, and smaller gaps between mean values and minimum values. More importantly, Algorithm \ref{alg:greedy_monotone2} achieves significantly higher $ss$ values in most settings, thus yielding better objective-diversity trade-offs than Algorithm \ref{alg:greedy_monotone}. This indicates benefits of limiting common elements by controlling their representation in the output, as they do not contribute to diversity.

Interestingly, increasing the output size $r$ only seems to affect the objective values, as the relative diversity values are virtually the same across all settings. We suspect that this might change for more complex matroids. Incidentally, the impacts of $r$ on objective values from Algorithm \ref{alg:greedy_monotone2} seem minimal outside of edge cases (i.e. $l=1$).
\section{Conclusion}\label{sec:conclusion}
The diverse solutions problem is a challenging extension to optimization problems that is of practical and theoretical interests. In this work, we considered the problem of finding diverse solutions to maximizing monotone submodular functions over a matroid. To address the difficulty in finding multiple high-quality solutions, we exploited submodularity with two simple greedy algorithms, equipped with objective-diversity trade-off adjusting parameters. Theoretical guarantees by these algorithms were given in both objective values and diversity, as functions of their respective parameters. Our experimental investigation with maximum vertex coverage instances demonstrates strong empirical performances from these algorithms despite their simplicity. 

\appendix


\section{On Algorithm \ref{alg:greedy_monotone2} Under Matroid Intersection}
We can generalize Theorem \ref{result:greedy_monotone2_matroid} and \ref{result:greedy_monotone2_matroid_P} to matroid intersection setting by extending the arguments. A set $x\in V$ satisfies the matroid intersection constraint defined by matroids $M_i=(V,\mathcal{I}_i)$ for $i=1,\ldots,k$ if $x\in\bigcap_{i=1}^k\mathcal{I}_i$. Let $W_{i,j}=cl_{M_j}\left(x^{(i-1)}\right)$, $U_{i,j}=V_i\cup W_{i,j}\setminus x^{(i-1)}$, $U_i=\bigcup_{j=1}^kU_{i,j}$ and $r_j(\cdot)=r_{M_j}(\cdot)$. We remark that matroid intersection is downward-closed, meaning all subsets of a feasible set are feasible.
\begin{theoremE}[][category=greedymonotone2kmatroid,end]\label{result:greedy_monotone2_kmatroid}
Given monotone submodular $f$ over $2^V$, $k\geq1$ matroids $M_i=(V,\mathcal{I}_i)$ such that $\mathcal{I}=\bigcap_j\mathcal{I}_j\supset\{\emptyset\}$, integers $r\geq2$ and $l\in[1,r)$, Algorithm \ref{alg:greedy_monotone2} run with inputs $f$, $\mathcal{I}$, $r$, $l$ outputs $P$ such that
\begin{align*}
\min\left\lbrace\frac{r-1}{l}+k+1,\max_{z\in\mathcal{I}}|z|\right\rbrace\min_{x\in P}f(x)\geq\max_{y\in\mathcal{I}}f(y).
\end{align*}
This bound is nearly tight for all $r\geq2$, $l\in[1,r)$, and $k\geq1$.
\end{theoremE}
\begin{proof}
For any $x\in P$, we have that since $y\in\mathcal{I}_j$ for all $j=1,\ldots,k$, we can use Lemma \ref{lemma:matroid_freedom_limit} for each of the $k$ matroids, and Lemma \ref{lemma:greedy_monotone2_discarded_elem} to derive for all $i=1,\ldots,|x|+1$
\begin{align*}
\left|y\cap U_i\right|&\leq\left|y\cap V_i\setminus x^{(i-1)}\right|+\sum_{j=1}^k\left|y\cap W_{i,j}\setminus x^{(i-1)}\right|\leq\left|y\cap V_i\setminus x^{(i-1)}\right|+\sum_{j=1}^kr_j\left(W_{i,j}\setminus x^{(i-1)}\right)\leq[\left(r-1\right)/l+k](i-1)
\end{align*}
Applying Lemma \ref{lemma:greedy_monotone2_ratio} yields the claim.

For tightness, let
\begin{align*}
f(x)&=\max\left\lbrace x_1,\left(1-\frac{\epsilon x_{l+1}}{k}\right)\right\rbrace+\sum_{i=2}^l\max\{x_i,x_{i+l}\}+\left(1-\frac{\epsilon x_1}{klr}\right)\sum_{i=2l+1}^{(k+2)l}x_i\\&+\sum_{i=(k+2)l+1}^{(k+2)l+r-1}x_i-\frac{\epsilon x_1}{klr}\left(\sum_{i=2}^{l}\max\{x_i,x_{i+l}\}\right)\left(\sum_{i=3l+1}^{3l+r-1}x_i\right),
\end{align*}
observe $f$ is monotone submodular for any $\epsilon\in[0,1]$. For $i=1,\ldots,k$, let matroid $M_i=(V,\mathcal{I}_i)$ be where $\mathcal{I}_i=\{x\subseteq V:\sum_{j=1}^l(x_j+x_{j+(i+1)l})\leq l\}$. Assuming w.l.o.g. that $\epsilon>0$, $v^*=1$, it can be the case that there is one solution $x'$ returned by Algorithm \ref{alg:greedy_monotone2} under $M$ whose second element is in $\{2,\ldots,l\}$ while the rest's are contained in $\{(k+2)l+1,\ldots,(k+2)l+r-1\}$ due to the fourth tie-breaking rule. After adding the $l$-th element to each solution, we can assume that $x'^{(l)}=\{1,\ldots,l\}$, meaning any element available to add to $x'$ at this point induces a marginal gain less than $1$. Since other solutions currently do not contain elements in $\{2,\ldots,(k+2)l\}$, and elements in $\{(k+2)l+1,\ldots,(k+2)l+r-1\}$ can still be added to them, inducing marginal gains $1$, the algorithm prioritizes them due to the second tie-breaking rule. Note that elements in $\{l+1,\ldots,(k+2)l\}$ are ignored in the next $r-1$ iterations since they induce marginal gains less than $1$ with $\epsilon>0$. When finally adding the $l+1$-th element to $x'$, $n_v(P)=l$ for all $v\in\{(k+2)l+1,\ldots,(k+2)l+r-1\}$, and elements in $\{2l+1,\ldots,(k+2)l\}$ cannot be considered, so $x'$ enjoys no further marginal gain, resulting in $f(x')=l$. We have $y=\{l+1,\ldots,(k+2)l+r-1\}\in\bigcap_{i=1}^k\mathcal{I}_k$ and $f(y)=r+l-1+k(l-\epsilon/k)=[(r-1)/l+k+1]f(x')-\epsilon$.
\end{proof}
\begin{theoremE}[][category=greedymonotone2kmatroidP,end]\label{result:greedy_monotone2_kmatroid_P}
Given monotone submodular $f$ over $2^V$, integers $r\geq2$, $l\in[1,r)$, $k\geq1$ matroids $M_i=(V,\mathcal{I}_i)$ where $\mathcal{I}=\bigcap_j\mathcal{I}_j\supset\{\emptyset\}$, and $Y\in\mathcal{I}^*$ such that $m=\max_{v\in V}n_v(Y)$, Algorithm \ref{alg:greedy_monotone2}, run with inputs $f$, $\mathcal{I}$, $r$, $l$ outputs $P$ such that
\begin{align*}
\min\left\lbrace\frac{m(r-1)}{l}+(k+1)|Y|,\sum_{y\in Y}|y|\right\rbrace\min_{x\in P}f(x)\geq\sum_{y\in Y}f(y).
\end{align*}
This bound is nearly tight for all $r\geq2$, $l\in[1,r)$, $k\geq1$, size of $Y$ and $m\in[1,|Y|]$.
\end{theoremE}
\begin{proof}
For any $x\in P$, we have $|V_i\setminus x^{(i-1)}|\leq (r-1)(i-1)/l$ for $i=1,\ldots,|x|+1$, so by the definition of $m$, $\sum_{v\in V_i\setminus x^{(i-1)}}n(Y)\leq m(r-1)(i-1)/l$. From the proof of Theorem \ref{result:greedy_monotone2_kmatroid}, we have for $i=1,\ldots,|x|+1$,
\[\sum_{j=1}^k\sum_{v\in U_i}n(Y)\leq\sum_{v\in V_i\setminus x^{(i-1)}}n(Y)+\sum_{y\in Y}\sum_{j=1}^k\left|y\cap W_{i,j}\setminus x^{(i-1)}\right|\leq\left[\frac{m(r-1)}{l}+k|Y|\right](i-1).\]
Applying Lemma \ref{lemma:greedy_monotone2_ratio} yields the claim.

For tightness, we fix $q\geq1$ and $m\in[1,q]$, let $w=\lceil q/m\rceil$, $o=[w(k+1)+1]l$ and
\begin{align*}
f(x)&=\max\left\lbrace x_1,\left(1-\frac{\epsilon z_{l+1}}{qk}\right)\right\rbrace+\sum_{i=2}^l\max\{x_i,z_{i+l}\}+\left(1-\frac{\epsilon x_1}{qklr}\right)\sum_{i=(w+1)l+1}^{(w+2)l}s_i\\&+\sum_{i=o+1}^{o+r-1}x_i-\frac{\epsilon x_1}{qklr}\left(\sum_{i=2}^{l}\max\{x_i,z_{i+l}\}\right)\left(\sum_{i=o+1}^{o+r-1}x_i\right),
\end{align*}
where $z_i=\max\{x_{jl+i}\}_{j=0}^{w-1}$ and $s_i=\sum_{j=0}^{k-1}z_{wlj+i}$, $f$ is monotone submodular for any $\epsilon\in[0,q]$. For $i=1,\ldots,k$, let matroid $M_i=(V,\mathcal{I}_i)$ be where $\mathcal{I}_i=\{x\subseteq V:\sum_{j=1}^lx_j+\sum_{j=1}^{wl}x_{j+(iw+1)l}\leq l\}$. Assuming w.l.o.g. that $\epsilon>0$, $v^*=1$, it can be the case that there is one solution $x'$ returned by Algorithm \ref{alg:greedy_monotone2} under $M$ whose second element is in $\{2,\ldots,l\}$ while the rest's are contained in $\{o+1,\ldots,o+r-1\}$ due to the fourth tie-breaking rule. After adding the $l$-th element to each solution, we can assume that $x'^{(l)}=\{1,\ldots,l\}$, meaning any element available to add to $x'$ at this point induces a marginal gain less than $1$. Since other solutions currently do not contain elements in $\{2,\ldots,o\}$, and elements in $\{o+1,\ldots,o+r-1\}$ can still be added to them, inducing marginal gains $1$, the algorithm prioritizes them due to the second tie-breaking rule. Note that elements in $\{l+1,\ldots,o\}$ are ignored in the next $r-1$ iterations since they induce marginal gains less than $1$ with $\epsilon>0$. When finally adding the $l+1$-th element to $x'$, $n_v(P)=l$ for all $v\in\{o+1,\ldots,o+r-1\}$, and elements in $\{(w+1)l+1,\ldots,o\}$ cannot be considered, so $x'$ enjoys no further marginal gain, resulting in $f(x')=l$. For $i=0,\ldots,q-1$, let $y_i=\bigcup_{h=1}^k\{(i\mod w)l+j,[(i\mod w)+hw]l+j\}_{j=l+1}^{2l}$, we have $y_i\cup\{o+1,\ldots,o+r-1\}\in\bigcap_{i=1}^k\mathcal{I}_k$. Let $Y=\left\lbrace y_i\cup\{o+1,\ldots,o+r-1\}\right\rbrace_{i=0}^{m-1}\cup\left\lbrace y_i\right\rbrace_{i=m}^{q-1}$, we have $m=\max_{v\in V}n_v(Y)$, and $\sum_{y\in Y}f(y)=m(r-1)+ql+qk(l-\epsilon/(qk))=[m(r-1)/l+(k+1)q]f(x')-\epsilon$.
\end{proof}
\section*{Acknowledgements}
This work was supported by the Australian Research Council through grants DP190103894 and FT200100536.
\bibliographystyle{ieeetr}
{\bibliography{refs}}
\end{document}